\documentclass[aps,prd,twocolumn,superscriptaddress,showpacs,preprintnumbers,amsmath,amssymb]{revtex4}

\usepackage{graphicx} 
\usepackage{dcolumn}  
\usepackage{amsmath} 

\def\myspecial#1{}                   

\begin{document}

\title{ \quad\\[0.5cm]   \boldmath Measurements of $CP$ Violation in 
     $B^0 \to D^{*-}\pi^+$ and $B^0 \to D^- \pi^+$ Decays
}

\affiliation{Budker Institute of Nuclear Physics, Novosibirsk}
\affiliation{Chiba University, Chiba}
\affiliation{Chonnam National University, Kwangju}
\affiliation{University of Cincinnati, Cincinnati, Ohio 45221}
\affiliation{University of Hawaii, Honolulu, Hawaii 96822}
\affiliation{University of Illinois at Urbana-Champaign, Urbana, Illinois 61801}
\affiliation{High Energy Accelerator Research Organization (KEK), Tsukuba}
\affiliation{Institute of High Energy Physics, Chinese Academy of Sciences, Beijing}
\affiliation{Institute of High Energy Physics, Vienna}
\affiliation{Institute of High Energy Physics, Protvino}
\affiliation{Institute for Theoretical and Experimental Physics, Moscow}
\affiliation{J. Stefan Institute, Ljubljana}
\affiliation{Kanagawa University, Yokohama}
\affiliation{Korea University, Seoul}
\affiliation{Kyungpook National University, Taegu}
\affiliation{Swiss Federal Institute of Technology of Lausanne, EPFL, Lausanne}
\affiliation{University of Ljubljana, Ljubljana}
\affiliation{University of Maribor, Maribor}
\affiliation{University of Melbourne, Victoria}
\affiliation{Nagoya University, Nagoya}
\affiliation{Nara Women's University, Nara}
\affiliation{National Central University, Chung-li}
\affiliation{National United University, Miao Li}
\affiliation{Department of Physics, National Taiwan University, Taipei}
\affiliation{H. Niewodniczanski Institute of Nuclear Physics, Krakow}
\affiliation{Nippon Dental University, Niigata}
\affiliation{Niigata University, Niigata}
\affiliation{Nova Gorica Polytechnic, Nova Gorica}
\affiliation{Osaka City University, Osaka}
\affiliation{Osaka University, Osaka}
\affiliation{Panjab University, Chandigarh}
\affiliation{Peking University, Beijing}
\affiliation{Princeton University, Princeton, New Jersey 08544}
\affiliation{RIKEN BNL Research Center, Upton, New York 11973}
\affiliation{Saga University, Saga}
\affiliation{University of Science and Technology of China, Hefei}
\affiliation{Seoul National University, Seoul}
\affiliation{Shinshu University, Nagano}
\affiliation{Sungkyunkwan University, Suwon}
\affiliation{University of Sydney, Sydney NSW}
\affiliation{Tata Institute of Fundamental Research, Bombay}
\affiliation{Toho University, Funabashi}
\affiliation{Tohoku Gakuin University, Tagajo}
\affiliation{Tohoku University, Sendai}
\affiliation{Department of Physics, University of Tokyo, Tokyo}
\affiliation{Tokyo Institute of Technology, Tokyo}
\affiliation{Tokyo Metropolitan University, Tokyo}
\affiliation{Tokyo University of Agriculture and Technology, Tokyo}
\affiliation{University of Tsukuba, Tsukuba}
\affiliation{Virginia Polytechnic Institute and State University, Blacksburg, Virginia 24061}
\affiliation{Yonsei University, Seoul}
   \author{F.~J.~Ronga}\affiliation{High Energy Accelerator Research Organization (KEK), Tsukuba} 
   \author{T.~R.~Sarangi}\affiliation{High Energy Accelerator Research Organization (KEK), Tsukuba} 
   \author{K.~Abe}\affiliation{High Energy Accelerator Research Organization (KEK), Tsukuba} 
   \author{I.~Adachi}\affiliation{High Energy Accelerator Research Organization (KEK), Tsukuba} 
   \author{H.~Aihara}\affiliation{Department of Physics, University of Tokyo, Tokyo} 
   \author{D.~Anipko}\affiliation{Budker Institute of Nuclear Physics, Novosibirsk} 
   \author{K.~Arinstein}\affiliation{Budker Institute of Nuclear Physics, Novosibirsk} 
   \author{Y.~Asano}\affiliation{University of Tsukuba, Tsukuba} 
   \author{T.~Aushev}\affiliation{Institute for Theoretical and Experimental Physics, Moscow} 
   \author{A.~M.~Bakich}\affiliation{University of Sydney, Sydney NSW} 
% ...
\collaboration{The Belle Collaboration}
\noaffiliation

\myspecial{!userdict begin /bop-hook{gsave 300 50 translate 5 rotate
    /Times-Roman findfont 18 scalefont setfont
    0 0 moveto 0.70 setgray
    (\mySpecialText)
    show grestore}def end}

\begin{abstract}
  We report measurements of time dependent decay rates for 
  $B^0 \rightarrow D^{(*)-}\pi^+$ decays 
  and extraction of $CP$ violation parameters that depend on $\phi_3$. 
  Using fully reconstructed $D^{(*)}\pi$ events and partially
 reconstructed  $D^{*}\pi$ events from a data 
  sample that contains 386 million $B\overline{B}$ pairs that was 
  collected near the $\Upsilon(4S)$ resonance, with the Belle 
  detector at the KEKB asymmetric energy $e^+ e^-$ collider,
  we obtain the $CP$ violation parameters 
  $S^+ (D^{(*)}\pi)$ and $S^- (D^{(*)}\pi)$.   
  We obtain 
  $S^+ (D^* \pi) = 0.049 \pm 0.020(\mathrm{stat}) \pm 0.011(\mathrm{sys})$, 
  $S^- (D^* \pi) = 0.031 \pm 0.019(\mathrm{stat}) \pm 0.011(\mathrm{sys})$,
 and 
  $S^+ (D \pi) = 0.031 \pm 0.030(\mathrm{stat}) \pm 0.012(\mathrm{sys})$, 
  $S^- (D \pi) = 0.068 \pm 0.029(\mathrm{stat}) \pm 0.012(\mathrm{sys})$. 
These results are an indication of $CP$ violation in $B^0 \to
 D^{*-}\pi^+$ and $B^0 \to D^- \pi^+$ decays at the $2.5\,\sigma$
 and $2.2\,\sigma$ levels, respectively.
If we use the values of $R_{D^{(*)}\pi}$ that are
derived using assumptions of factorization and  SU(3) symmetry, the 
branching fraction measurements for the $D_s^{(*)} \pi$
modes, and lattice QCD calculations, we can restrict the allowed region
 of $|\sin (2\phi_1 + \phi_3)|$ to be above 0.44 and 0.52 at 68\%
 confidence level from 
the $D^* \pi$ and $D \pi$ modes, respectively.  
\end{abstract}


\maketitle

\tighten


\section{Introduction}

Within the Standard Model (SM), 
$CP$ violation arises due to a single phase
in the Cabibbo-Kobayashi-Maskawa (CKM) quark mixing matrix~\cite{KM}.
Precise measurements of CKM matrix parameters therefore
constrain the SM, and may reveal new sources of $CP$ violation.
Measurements of the time-dependent decay rates of 
$B^0 (\overline{B}{}^0) \to D^{(*)\mp}\pi^{\pm}$
provide a theoretically clean method for extracting 
$\sin(2\phi_1+\phi_3)$~\cite{dunietz}.
As shown in Fig.~\ref{fig:feynman},
these decays can be mediated by both
Cabibbo-favoured decay (CFD) and doubly-Cabibbo-suppressed decay (DCSD) 
amplitudes, 
$V_{cb}^*V_{ud}$ and $V_{ub}^*V_{cd}$, which have a relative weak phase 
$\phi_3$.
\begin{figure}[hbt]
\begin{minipage}{4.25cm}
\includegraphics[width=4.0cm,clip]{dpi_cfd.epsi}\\
(a) CFD\\
\end{minipage}
\begin{minipage}{4.25cm}
\includegraphics[width=4.0cm,clip]{dpi_dcsd.epsi}\\
(b) DCSD\\
\end{minipage}
    \caption{
      Diagrams for 
      (a) $B^0 \to D^{(*)-}\pi^+$ and 
      (b) $\overline{B}{}^0 \to D^{(*)-}\pi^+$.
      Those for $\overline{B}{}^0 \to D^{(*)+}\pi^-$ and 
      $B^0 \to D^{(*)+}\pi^-$
      can be obtained by charge conjugation.}
      \label{fig:feynman}
\end{figure}

The time-dependent decay rates are given by~\cite{fleischer}
\begin{eqnarray}
P(B^{0} &\to& D^{(*)+} \pi^-) = \frac{1}{8\tau_{B^0}} 
                    e^{-|\Delta t|/\tau_{B^0}} \nonumber \\ 
&&   \times \left[
      1 - C \cos (\Delta m \Delta t) - S^+ \sin (\Delta m \Delta t) 
    \right],  \nonumber \\
P(B^{0} &\to& D^{(*)-} \pi^+) = \frac{1}{8\tau_{B^0}}
                  e^{-|\Delta t|/\tau_{B^0}} \nonumber \\ 
&& \times    \left[
      1 + C \cos (\Delta m \Delta t) - S^- \sin (\Delta m \Delta t) 
    \right], \nonumber \\  
P(\overline{B}{}^0 &\to& D^{(*)+} \pi^-) =  \frac{1}{8\tau_{B^0}}  
                  e^{-|\Delta t|/\tau_{B^0}}\nonumber \\ 
&&  \times    \left[
      1 + C \cos (\Delta m \Delta t) + S^+ \sin (\Delta m \Delta t) 
    \right], \nonumber \\  
P(\overline{B}{}^0 &\to& D^{(*)-} \pi^+) = \frac{1}{8\tau_{B^0}}  
                  e^{-|\Delta t|/\tau_{B^0}}\nonumber \\ 
&&  \times     \left[
      1 - C \cos (\Delta m \Delta t) + S^- \sin (\Delta m \Delta t) 
    \right]. 
    \label{eq:evol}  
  \end{eqnarray}
Here $\Delta t$ is the difference between the time of the decay and the 
time that the flavour of the $B$ meson is tagged,  
$\tau_{B^0}$ is the average neutral $B$ meson lifetime, 
$\Delta m$ is the $B^0$-$\overline{B}{}^0$ mixing parameter, and 
$C = \left( 1 - R^2 \right) / \left( 1 + R^2 \right)$, 
where $R$ is the ratio of the magnitudes of the DCSD and CFD 
(we assume the magnitudes of both the CFD and  DCSD amplitudes are the
same for $B^0$ and $\overline{B}{}^0$ decays). 
The $CP$ violation parameters are given by
\begin{equation}
S^{\pm} = \frac{2 (-1)^L R \sin(2\phi_1+\phi_3 \pm \delta)}
               { \left( 1 + R^2 \right)},
\label{eq:spm}
\end{equation}
where $L$ is the orbital angular momentum of
the final state (1 for $D^* \pi$ and 0 for $D \pi$), and $\delta$ is 
the strong phase difference of the CFD and DCSD. The values of 
$R$ and $\delta$
are not necessarily the same for $D^* \pi$ and $D \pi$ final states, and
are denoted with subscripts, $D^* \pi$ and $D \pi$, in what follows.  

Although not measured yet, the value of $R$ is predicted to be about 
$0.02$~\cite{csr}. Therefore, 
we neglect terms of ${\cal O}\left( R^2 \right)$ 
(and hence take $C = 1$).  
There are theoretical arguments that the still unmeasured values
of $\delta$ for both $D^*\pi$ and $D \pi$ are small
~\cite{fleischer,wolfenstein}.

Due to the size of $R$, $CP$ violation is expected to be a small effect
in these decays. Therefore, a large event sample is needed in order to
obtain sufficient sensitivity. With this in mind, we employ a partial 
reconstruction technique~\cite{zheng} for the $D^* \pi$ analysis in
addition to the conventional full reconstruction method.
In this approach, the signal is distinguished from background on the basis of 
kinematics of the ``fast'' pion from the decay $B \to D^* \pi_f$,
and the ``slow'' pion from the decay $D^* \to D \pi_s$, alone;
no attempt is made to reconstruct the  $D$ meson from its decay products.
Background from continuum $e^+e^- \to q\overline{q} \ (q = u,d,s,c)$
events is reduced dramatically by requiring the presence 
of a high-momentum lepton in the event,
which also serves to tag the flavour of the associated $B$ in the event.

Results from our previous analyses using a $140\,\mathrm{fb}^{-1}$ data sample
containing 152 million $B \overline{B}$ pairs have been published for 
the full reconstruction method~\cite{belle_full} and the  partial
reconstruction method~\cite{belle_partial}. Results from similar
analyses by BaBar collaboration were also reported~\cite{babar_full,
babar_partial}. 
This study is a continuation of similar analyses with a substantially
increased data sample containing 386 million 
$B \overline{B}$ events, and several improvements in the analyses.

\section{Belle Detector}
The data was  collected with the Belle
detector~\cite{Belle} at the KEKB asymmetric energy electron-positron 
collider~\cite{KEKB} operating at the $\Upsilon$(4S) resonance. 
The Belle detector is a large-solid-angle magnetic 
spectrometer that consists of a silicon vertex detector (SVD), 
a 50-layer central drift chamber (CDC), an array of aerogel 
threshold Cherenkov counters (ACC), a barrel-like arrangement 
of time-of-flight scintillation counters (TOF), and an 
electromagnetic calorimeter (ECL) comprised of CsI(Tl) 
crystals located inside a superconducting solenoidal coil 
that provides a 1.5 T magnetic field. An iron flux-return 
located outside of the coil is instrumented to detect $K_L^0$ 
mesons and to identify muons (KLM). 
A sample containing 
152 million $B \overline{B}$ pairs was collected with 
a 2.0~cm radius beampipe and a 3-layer silicon vertex detector 
(SVD1), 
while a sample of 234 million $B \overline{B}$ pairs was 
collected with a 1.5~cm radius beampipe, a 4-layer silicon 
detector, and a small-cell inner drift chamber (SVD2)
~\cite{svd2}.

\section{Full Reconstruction Analysis}
\subsection{Signal Event Selection}
The selection of hadronic events is described elsewhere~\cite{hadsel}. 
For the $\overline{B}{}^0 \to D^{*+} \pi^-$ event selection, 
we use the decay chain $D^{*+}\to D^+\pi^0$ or $D^0\pi^+$ with 
subsequent decays of   
$D^+\to K^-\pi^+\pi^+$ and $D^0\to K^-\pi^+,~K^-\pi^+\pi^0,~ 
K^-\pi^+\pi^+\pi^-,~ K_S^0\pi^+\pi^- (K_S^0\to \pi^+ \pi^-)$.  
(Charge conjugate modes are implied throughout this Paper.)
All charged tracks except for the slow pions from $D^* \to D \pi$ 
decays are required to have a minimum of one hit (two hits) in 
the $r$-$\phi$ ($z$) coordinate of the vertex detector, where 
the $r$-$\phi$ plane is transverse to the positron beam
line that defines the $z$ axis. 
These requirements allow a precise determination of the production 
point.
To separate kaons from pions, we form a 
likelihood for each track, ${\cal L}_{K(\pi)}$. The kaon 
likelihood ratio, $P(K/\pi) = {\cal L}_K / ({\cal L}_K 
+ {\cal L}_{\pi})$, has values between 0 (likely to be a pion) 
and 1 (likely to be a kaon). We require charged kaons to 
satisfy $P(K/\pi)>0.3$, corresponding to about  95\% efficiency 
for detecting kaons and about  2\% probability for misidentifying 
pions as kaons. 
There is no such requirement for 
charged pions coming from $D$ decays. 

Neutral pions are formed from photon pairs with invariant
masses between $0.118\,\mathrm{GeV}/c^2$ and $0.150\,\mathrm{GeV}/c^2$.
The photon momenta are then recalculated with a $\pi^0$ mass constraint.

We require the invariant mass of $K_S^0 \to \pi^+ \pi^-$ candidates to 
be between $0.485\,\mathrm{GeV}/c^2$ and $0.510\,\mathrm{GeV}/c^2$ 
corresponding to $\pm 5\, \sigma$, where $\sigma$ is the Monte Carlo
(MC) determined invariant mass resolution. 
The radial impact parameter, which
is the distance of closest approach of the
candidate charged tracks to the event-dependent interaction point (IP) 
in the $r$-$\phi$ 
plane, is required to be larger than $0.02\,\mathrm{cm}$ for high momentum 
($>\,1.5 \,\mathrm{GeV}/c)$ $~K_S^0$ candidates and $0.03\,\mathrm{cm}$ 
for those with momentum less than $1.5\,\mathrm{GeV}/c$. 
Here the IP is determined for each set of 10,000 neighboring hadronic
events, and the $K_S^0$ momentum is given
in the $\Upsilon$(4S) rest frame (cms). 
The $\pi^+ ~\pi^-$
vertex is required to be displaced from the IP by a minimum 
transverse
distance of $0.22\,\mathrm{cm}$ for high momentum candidates and
$0.08\,\mathrm{cm}$ for
the remaining candidates. The mismatch in the $z$ direction
at the $K_S^0$ vertex point for the $\pi^+ ~\pi^-$ tracks must be less
than $2.4\,\mathrm{cm}$ for high momentum candidates and
$1.8\,\mathrm{cm}$  for the 
remaining candidates. The direction of the pion pair momentum must also 
agree with the direction of the vertex point from the IP to
within $0.03\,\mathrm{rad}$ for high momentum candidates and to within 
$0.1\,\mathrm{rad}$ for the remaining candidates.
A mass and vertex constraint is 
imposed when fitting the $K_S^0$ candidates.

For $D^0$ meson candidates, the invariant mass of the $D^0$ candidate is 
required to 
be within $\pm 20$, $\pm 30$, $\pm 20$, and 
$\pm 20\,\mathrm{MeV}/c^2$ for $K^-\pi^+,~ K^-\pi^+\pi^0,~ 
K^-\pi^+\pi^+\pi^-,~ K_S^0\pi^+\pi^-$ modes respectively,
while the invariant mass of the $D^+$ candidates should be within 
$\pm 18\,\mathrm{MeV}/c^2$ 
of the nominal $D^+$ mass corresponding to $\pm 4\,\sigma$. 
For the $D^0 \to K^- \pi^+ \pi^0$ mode, 
we further require the $\pi^0$ cms momentum to be greater than 
$200\,\mathrm{MeV}/c$. 
We use a mass- and vertex-constrained fit for $D$ candidates.

The $D^{*+}$ is reconstructed by combining either $D^0$ 
candidates with a slow $\pi^+$ meson, or $D^+$ candidates with a 
slow $\pi^0$ meson. 
The $D^*$ candidates are required to have a mass difference 
$\Delta M = M_{D\pi} - M_D$  within $\pm 3\,\mathrm{MeV}/c^2$
to $\pm 5\,\mathrm{MeV}/c^2$ of the nominal value depending on the decay mode.

We reconstruct a $B$ candidate by combining the $D^{(*)+}$ candidate with 
a $\pi^-$ candidate satisfying $P(K/\pi) < 0.8$, corresponding to more
than 90\% efficiency for detecting pions and less than 10\% probability
for misidentifying kaons as pions. 
We identify $B$ decays based on requirements on the energy difference 
$\Delta E \equiv \sum_i E_i - E_{\rm{beam}}$ and the  
beam-energy constrained mass 
$M_{\rm{bc}} \equiv \sqrt{E_{\rm{beam}}^2 - (\sum_i \vec{p}_i)^2}$, 
where $E_{\rm{beam}}$ is the beam energy, and $\vec{p}_i$ and $E_i$ are 
the momenta and energies of the daughters of the reconstructed $B$ 
meson candidate, all in the cms. 
We define a signal region in the 
$\Delta E$-$M_{\mathrm{bc}}$ plane of 
$5.27\,\mathrm{GeV}/c^2 < M_{\rm bc} < 5.29\,\mathrm{GeV}/c^2$ and 
$\left| \Delta E \right| < 0.045\,\mathrm{GeV}$, 
corresponding to about $\pm 3\sigma$ on each quantity.
If more than one $B$ candidate is found in the same event, 
we select the one with best $D$ vertex quality.
For the determination of background parameters, 
we use events in a sideband region defined by 
$M_{\rm bc} > 5.2\,\mathrm{GeV}/c^2$ and
$-0.14\,\mathrm{GeV} < \Delta E  < 0.20\,\mathrm{GeV}$, 
excluding the signal region. 

\begin{figure}[!tbh]
\begin{minipage}{4.25cm}
\includegraphics[width=4.25cm,clip]{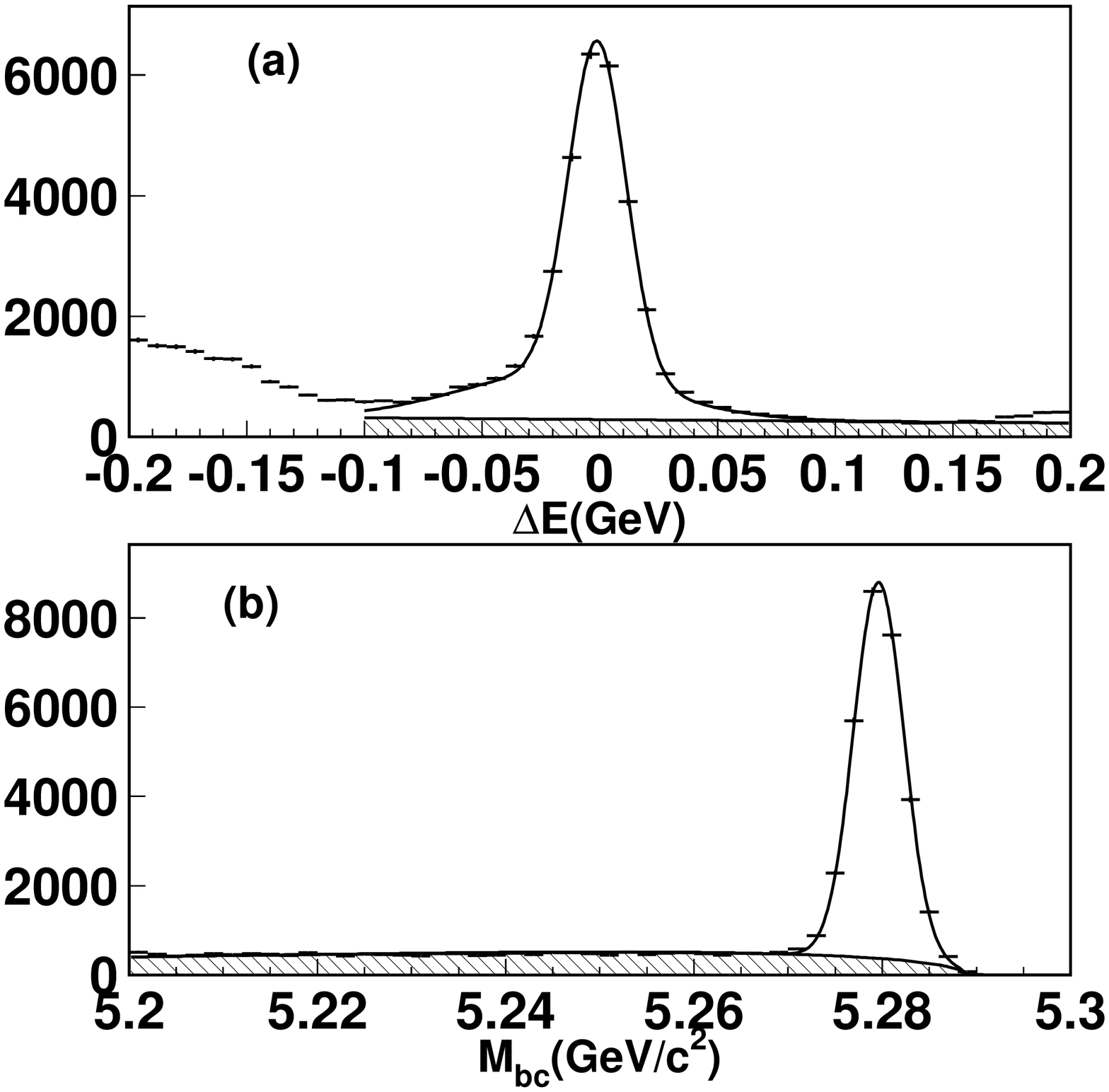}
\end{minipage}
\begin{minipage}{4.25cm}
\includegraphics[width=4.25cm,clip]{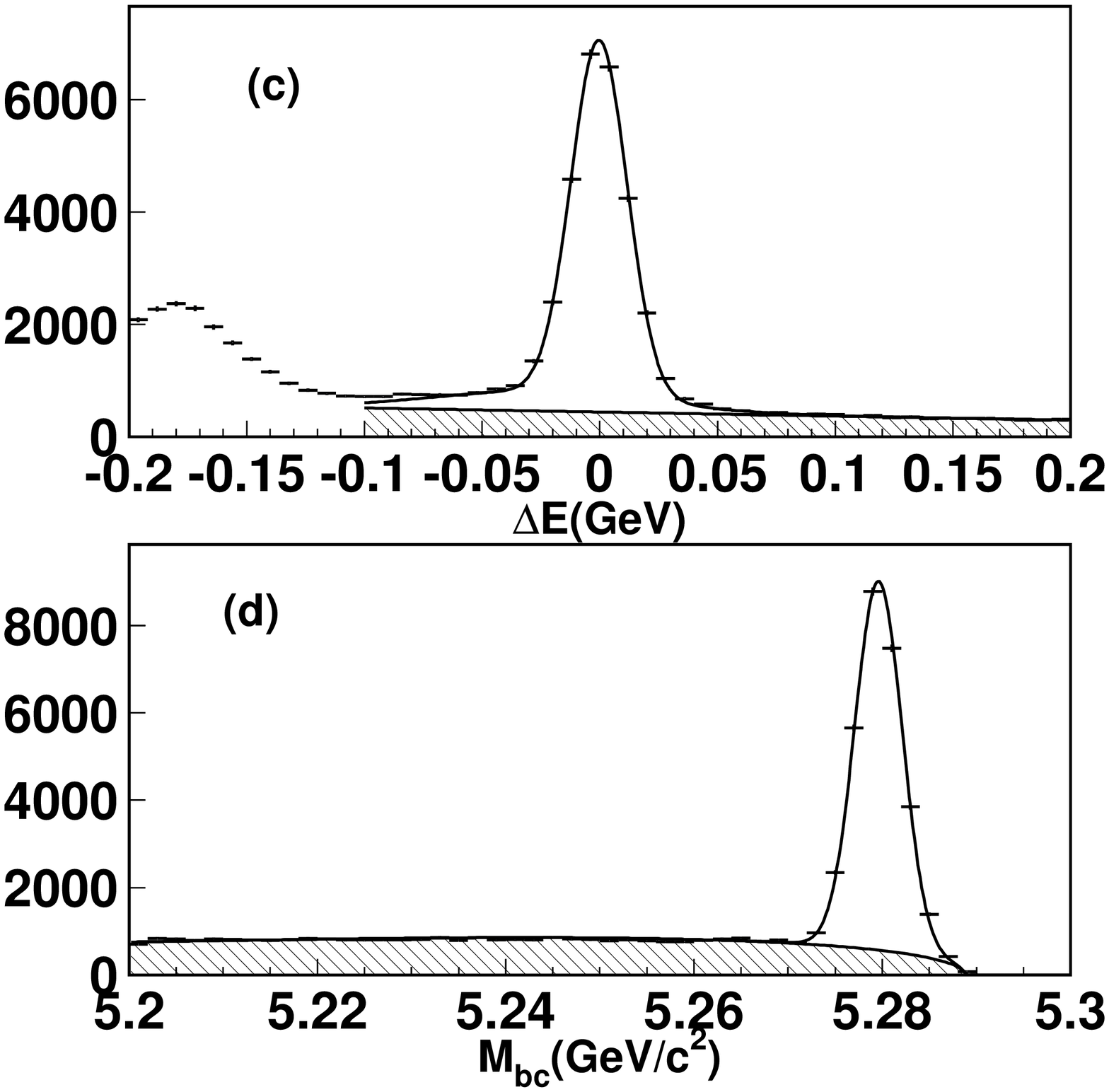}
\end{minipage}
    \caption{
     $\Delta E$ and $M_{\rm{bc}}$ distributions for $B^0 \to D^{*-}
 \pi^+$ candidates (a and b), and  $B^0 \to D^- \pi^+$ candidates 
    (c and d). Curves are the fit results. Hatched regions indicate 
the background components of the fits.
}
      \label{fig:dembc}
\end{figure}

The $\Delta E$ and $M_{\rm{bc}}$ distributions for $D^* \pi$ and $D \pi$
candidates are shown in Fig.~\ref{fig:dembc}.
A study using MC events indicates the presence of ``peaking
background'' that peaks in the signal $\Delta E$-$M_{\mathrm{bc}}$
region and amounts to 1.7\% (0.7\%) of the $D^* \pi$ ($D
\pi$) candidates, respectively. We treat this contribution as a part of 
the signal, and assign a systematic error to account for this.

\subsection{Flavour Tagging}
Charged leptons, pions, and kaons that are not associated with the
reconstructed $D^{(*)} \pi$ decays are used to identify the flavour of
the accompanying $B$ meson. 
The algorithm~\cite{belle_sin2phi1} produces two parameters for each event,
$q$ and $r$, where $q=+1$($q=-1$) for a $B^0$ ($\overline{B}{}^0$) meson and 
$r$ is a quality factor ranging from 0 for no flavour
discrimination to 1 for unambiguous flavour assignment.  
The algorithm uses the kinematical variables of the event 
and compares to those from a large number of MC events, 
and is used only to sort the data into six intervals of $r$  
according to estimated flavour purity.
More than 99.5\% of the events are assigned non-zero values of $r$.       

\subsection{Vertex Measurement}
                                                
The decay vertex of the $B \to D^{(*)} \pi$ candidate is fitted 
using the track information of the $D$ and $\pi$ (except the slow 
$\pi$ from $D^{*}$ decay). 
For the decay vertex of the tagging $B$ meson,          
the remaining well-reconstructed tracks in the event are used. 
Tracks that are consistent with $K^0_S$ decay are rejected.  
We impose the additional requirement that both signal-side and
tagging-side vertices be consistent with the run-dependent IP profile. 
A study using a MC event sample shows that 
the nominal $\chi^2$ of the vertex reconstruction is strongly correlated 
with the distance from the IP to the reconstructed vertex.                     
To avoid possible bias in the event selection due to this correlation, 
we introduce a quantity that only depends on the $z$ coordinate quantities;   
\begin{equation} 
\xi \equiv \frac{1}{2n}\sum^n_i\left|\frac{z^{\rm{after}}_i - 
z^{\rm{before}}_i}{\epsilon^{\rm{before}}_i}\right|^2,  
\label{eq:xi} 
\end{equation}
where $n$ is the number of tracks,  $z_i$ are the $z$ vertices of the $i$-th 
track before and after the  vertex fit, and $\epsilon^{\rm {before}}_i$ 
is the measurement error of the $z$ before the vertex fit. This quantity   
is calculated for the signal- and tagging-side separately, only for         
cases with multiple tracks. We require $\xi < 100$ to eliminate badly 
reconstructed vertices, which amount to 4\% (2\%) of the signal-side and 
1\% (1\%) of the tagging-side in the
 $D^*\pi$ and $D \pi$ modes, respectively.

The proper-time difference between the vertices $z_{\rm rec}$
and $z_{\rm tag}$ (measured along the beam line) of the fully reconstructed 
$B$ candidate and the tagging $B$ meson, respectively, is calculated as  
\begin{equation}
\Delta t = (z_{\mathrm{rec}}-z_{\mathrm{tag}})/\beta \gamma c,
\label{eq:dtvsdz}
\end{equation}
where $\beta \gamma = 0.425$ is the
Lorentz boost factor of the centre of mass frame at KEKB. 
After application of the event selection criteria and the  requirement that 
both $B$ mesons have well defined vertices and 
$\left| \Delta t \right| < 70\,\mathrm{ps}$ ($\sim 45\, \tau_{B^0}$),  
31491 and 31725 candidates remain in the $D^* \pi$ and $D\pi$ modes, 
respectively. The signal fractions of the candidates, which vary for
different $r$ bins, are 89\% for $D^* \pi$ and 83\% for $D\pi$. 

\subsection{\boldmath $\Delta t$ Fit}
Unbinned maximum likelihood fits to the four time dependent decay
rates are performed to extract
$S^+$ and $S^-$. 
We minimize $-2 \sum_{i} \ln {\cal L}_i$ where the likelihood for
the $i$-th event is given by
\begin{equation} 
 {\cal L}_i = (1 - f_{\rm ol})
  \left[
    f_{\rm sig} P_{\rm sig} 
    + (1-f_{\rm sig}) P_{\rm bkg}
  \right]  + f_{\rm ol} P_{\rm ol}.
\end{equation}
The signal fraction $f_{\mathrm{sig}}$ is 
determined from the  ($\Delta E$-$M_\mathrm{bc}$) value of each event. 
The signal distribution is the product of the sum of two Gaussians in 
$\Delta E$ and a Gaussian in $M_{\rm bc}$; that for the background is 
the product of a first order polynomial in $\Delta E$ and an ARGUS 
function~\cite{argus} in $M_{\rm bc}$.

The signal $\Delta t$ distribution is given by 
\begin{eqnarray}
P(q=-1, D^{(*)\pm} \pi^{\mp}) &=&    
  (1- w_-) P(B^0 \to D^{(*)\pm} \pi^{\mp}) \nonumber \\
  &+& 
       w_+  P(\overline{B}{}^0 \to D^{(*)\pm} \pi^{\mp}) 
  \label{eq:pdf1}
\end{eqnarray}
for the $q=-1$ sample, and 
\begin{eqnarray}
P_(q=+1, D^{(*)\pm} \pi^{\mp}) &=& 
   (1- w_+) P(\overline{B}{}^0 \to D^{(*)\pm} \pi^{\mp}) \nonumber \\
   &+& 
       w_-  P({B^0} \to D^{(*)\pm} \pi^{\mp})  
   \label{eq:pdf2}
\end{eqnarray}
for the $q=+1$ sample.
Here  $w_+$ and $w_-$ are respectively the probabilities to incorrectly 
assign the flavour of tagging $B^0$ and $\overline{B}{}^0$ mesons 
(wrong tag fractions), 
and the decay rates are given by Eq.~\ref{eq:evol}. 

The corresponding background distribution is parameterized as a sum of a 
$\delta$-function component and an exponential component with lifetime 
$\tau_{\rm bkg}$:
\begin{equation}
  \delta(\Delta t -\mu_{\rm bkg}^{\delta}) + 
  \frac{(1-f_{\rm bkg}^{\delta})}{2\tau_{\rm bkg}}
  e^{-\left| \Delta  t -\mu_{\rm bkg}^\tau \right|/{\tau_{\rm bkg}}} ,
        \label{eq:bkgpdf}
\end{equation}
where $f_{\rm bkg}^{\delta}$ is the fraction of events 
contained in the $\delta$-function, 
$\mu_{\rm bkg}^{\delta}$ and $\mu_{\rm bkg}^\tau$
are the mean values of $\Delta t $ in the 
$\delta$-function and  exponential components, respectively.
These parameters are determined separately from a fit to the 
$\Delta t$                
distribution in the $\Delta E$-$M_\mathrm{bc}$ sideband data                 
for the $D^* \pi$ and $D \pi$ data samples, SVD1 and              
SVD2 data, and cases where the vertices are reconstructed using              
single track and multiple tracks. Typically, values of the background
$\Delta t$ parameters are:              
$f_\mathrm{bkg}^\delta \approx 0.4$, 
$\tau_\mathrm{bkg} \approx 1.1\,\mathrm{ps}$, 
$\mu_\mathrm{bkg}^\tau \approx -0.1\,\mathrm{ps}$,          
and $\mu_\mathrm{bkg}^\delta \approx 0$.

A small number of events have poorly reconstructed vertices resulting in
very broad $\Delta t$ distributions. 
We account for this outlier contribution by
adding a Gaussian component $P_{\rm ol}$ with a width and fraction
determined from the $B$ lifetime analysis~\cite{vertex_nim}.

\subsection{\boldmath $\Delta t$ Resolution}
The  probability density functions (PDF) for the signal and background
must be convolved with corresponding $\Delta t$ resolution functions,  
that are determined on an event-by-event basis using the 
estimated uncertainties on the $z$ vertex positions~\cite{vertex_nim,
response}, in order to be compared with the data. 
     The signal resolution function is a convolution of
three contributions: resolution functions for vertex reconstruction, 
smearing due to non-primary tracks ($K_S^0$ and charm daughters),
 and the kinematic approximation that the $B$ mesons are at rest in 
the cms.
The resolution function is described in detail elsewhere~\cite{vertex_nim}.

The background resolution function is parametrized as a
sum of two Gaussians where different values are used for the parameters
depending on whether or not both vertices are reconstructed with
multiple tracks. 
These parameters are determined from the $\Delta E$-$M_\mathrm{bc}$
sideband data.

\subsection{\boldmath Tagging Side $CP$ Violation Effect}
While the tagging side should have no asymmetry if the flavour is tagged by 
primary leptons since semileptonic decays are flavour-specific
processes, it is possible to introduce a small 
asymmetry when daughter particles of hadronic 
decays such as $D^{(*)} \pi$ are used for the flavour tagging,
due to the same $CP$ violating effect that is the subject of 
this paper ~\cite{tagsidecpv}. This effect is taken into account by 
replacing the $S^\pm$ parameters in Eq.~\ref{eq:evol} 
by 
\begin{eqnarray}
B^0 \to D^{(*)+}\pi^- &:& S^+ \to (S^+ - S^+_\mathrm{tag}),
                                                     \nonumber \\ 
B^0 \to D^{(*)-}\pi^+ &:& S^- \to (S^- + S^+_\mathrm{tag}),
                                                     \nonumber \\
\overline{B}{}^0 \to D^{(*)+}\pi^- &:& S^+ \to (S^+ + S^-_\mathrm{tag}),
                                                     \nonumber \\ 
\overline{B}{}^0 \to D^{(*)-}\pi^+ &:& S^- \to (S^- - S^-_\mathrm{tag}), 
\label{eqntag}
\end{eqnarray}
respectively.
Here $S^+_\mathrm{tag}$ and $S^-_\mathrm{tag}$   
represent the $CP$ violation effect on the flavour tagging side due to the 
presence of $B^0 \to \overline{D}~\overline{X}$ and 
$\overline{B}{}^0 \to D X$ amplitudes, respectively.
Note that unlike the $S^\pm$ parameters, which are defined rigorously 
in terms of 
$B^0 \to D^{(*)\mp} \pi^{\pm}$ and $\overline{B}{}^0 \to D^{(*)\pm} \pi^{\mp}$ 
amplitudes, $S^\pm _\mathrm{tag}$ 
are effective quantities that include effects of the fraction of 
$B \to DX (\overline{D}~\overline{X})$ components in the tagging $B$
decays and the  
subsequent behaviour of $D (\overline{D})$ mesons. 
Therefore, these quantities must be determined experimentally. 

The values of $S^\pm _\mathrm{tag}$ are determined in each 
$r$ bin by fitting the $\Delta t$ distributions 
of a $B \to D^*l\nu$ control sample~\cite{belle_sin2phi1} 
using the signal PDFs of Eqs.~\ref{eq:pdf1} and \ref{eq:pdf2} 
and setting
$S^\pm$ to zero. 
Since the $D^*l\nu$ final states have specific flavour, any observable 
asymmetry must originate from the tagging side.  
The results for each $r$ bin are listed in Table~\ref{tab:s-tag}. 
The errors listed here are statistical only.    
The result for the combined $r$ bins are
$S^+_\mathrm{tag} = -0.002 \pm 0.009 \pm 0.006$ and
$S^-_\mathrm{tag} = 0.017 \pm 0.009 \pm 0.006$, where the second errors
are systematic.

\begin{table}[htb]
\caption{
  Parameters describing tagging side $CP$ violation effect that are
 determined from the $D^* l \nu$ data sample.
}
\label{tab:s-tag}
\begin{tabular}{ccccc}
\hline \hline
$r$ & $S^+_\mathrm{tag}$ & $S^-_\mathrm{tag}$ \\
\hline
 0.000~--~0.250~~~~ & $-0.058 \pm 0.130$~~~~ & $+0.060 \pm 0.130$ \\
 0.250~--~0.500~~~~ & $+0.001 \pm 0.040$~~~~ & $+0.018 \pm 0.040$ \\
 0.500~--~0.625~~~~ & $+0.027 \pm 0.032$~~~~ & $-0.030 \pm 0.032$ \\
 0.625~--~0.750~~~~ & $+0.026 \pm 0.025$~~~~ & $+0.022 \pm 0.025$ \\
 0.750~--~0.875~~~~ & $-0.011 \pm 0.025$~~~~ & $+0.027 \pm 0.025$ \\
 0.875~--~1.000~~~~ & $-0.005 \pm 0.014$~~~~ & $+0.024 \pm 0.014$ \\
\hline \hline
\end{tabular}
\end{table}

\subsection{Fit Result}
\begin{figure}[!t]
\begin{minipage}{4.25cm}
\includegraphics[width=4.25cm,clip]{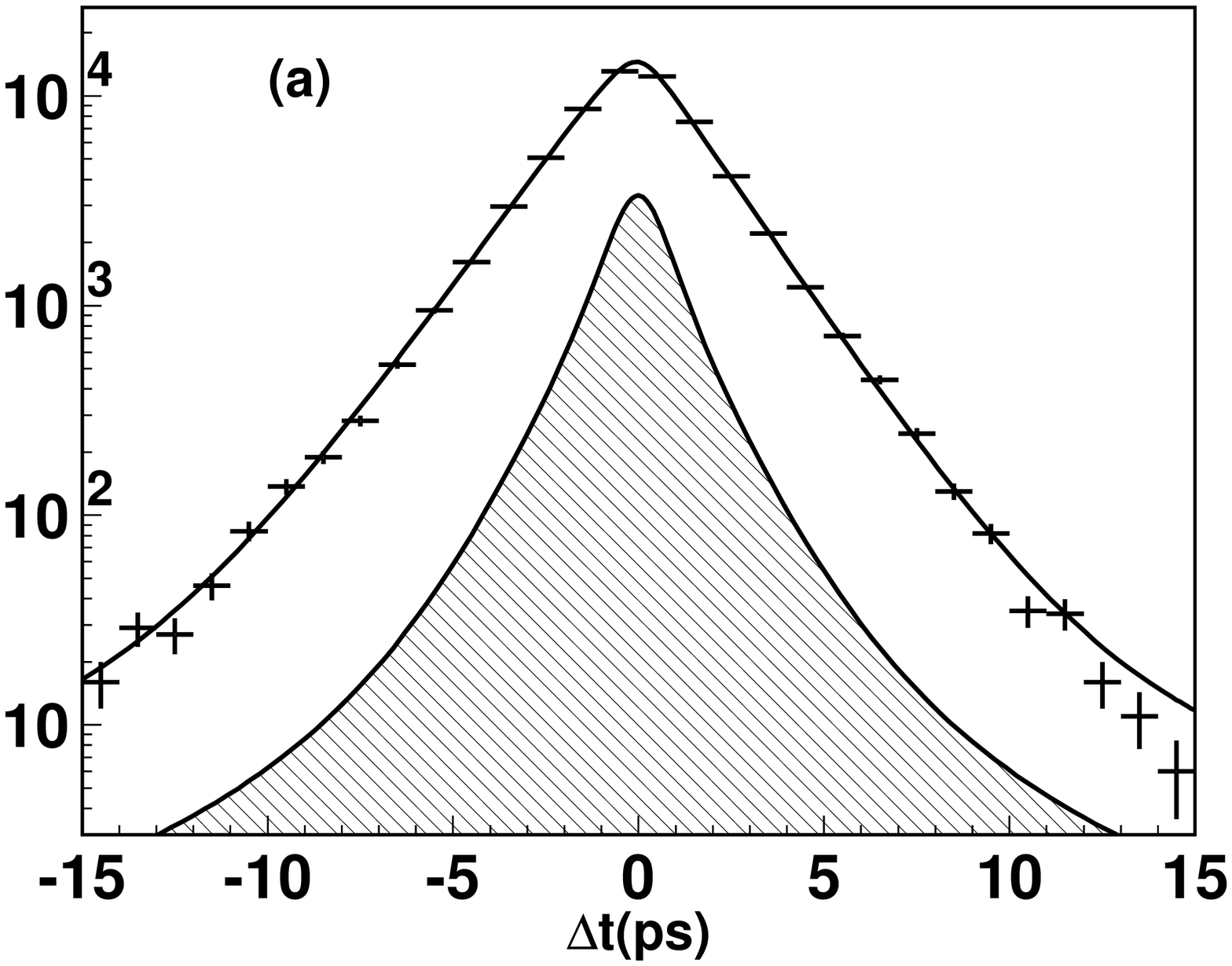}
\end{minipage}
\begin{minipage}{4.25cm}
\includegraphics[width=4.25cm,clip]{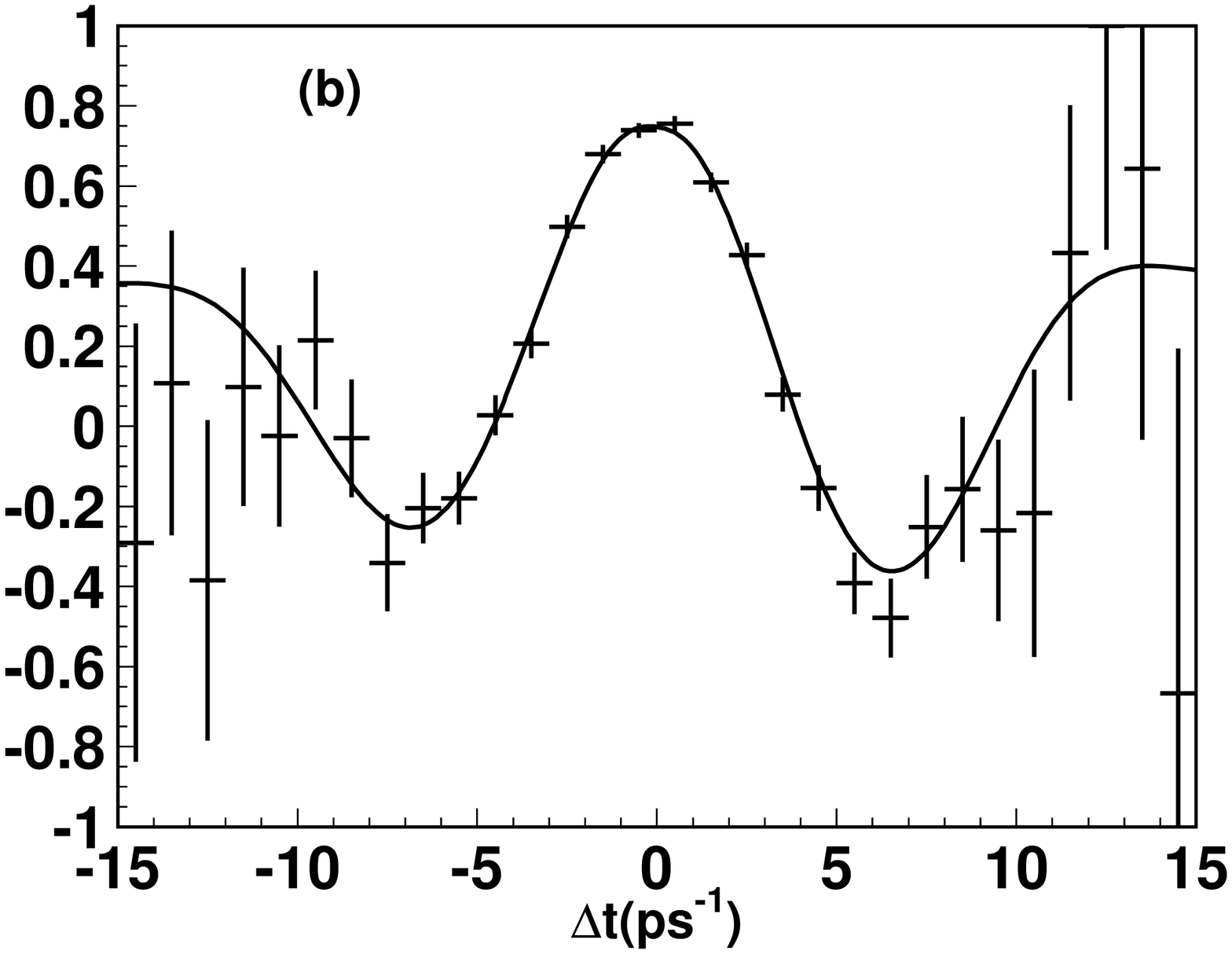}
\end{minipage}
    \caption{
      (a) $\Delta t$ distribution for the signal candidates 
 when all four signal categories are combined. 
      (b) $\mathrm{(OF-SF)/(OF+SF)}$ asymmetries. Curves  are the fit results. 
     The hatched region in (a) indicates the background.}
      \label{fig:mixing}
\end{figure}

The procedures for $\Delta t$ determination and flavour tagging are
tested by extracting $\tau_{B^0}$ and $\Delta m$. 
When all four signal categories in Eq.~\ref{eq:evol} are combined, 
the signal $\Delta t$ distribution reduces to an exponential lifetime 
distribution as shown in Fig.~\ref{fig:mixing}(a).  
We do a simultaneous fit to the SVD1 and SVD2 samples by combining 
the $D^*\pi$ and $D\pi$ candidate events and obtain 
$\tau_{B^0} = 1.532 \pm 0.013 \ {\rm ps}$, where the error 
is statistical only, in good agreement with the
world average of $1.536 \pm 0.014~\mathrm{ps}$~\cite{PDG}. 
Combining the two CFD-dominant modes (denoted as OF because the
signal-side and tagging-side have opposite $B$ flavour) and the two 
mixing-dominant modes (denoted as SF because the signal-side and
tagging-side have same $B$ flavour) and ignoring the
$CP$ violating terms, an asymmetry $\mathrm{(OF - SF)/(OF + SF)}$ behaves as 
$\cos (\Delta m \Delta t)$ as shown in Fig.~\ref{fig:mixing}(b). 
A similar fit to the one used for the lifetime determination gives 
$\Delta m = 0.494 \pm 0.007 \ {\rm ps}^{-1}$, where the error is
statistical only, also in good agreement with 
the world average $0.502 \pm 0.007~\mathrm{ps^{-1}}$~\cite{PDG}.
This fit is also used to determine the signal $\Delta t$ resolution 
parameters and wrong tag fractions $w_+$ and $w_-$ in each $r$ bin for 
both SVD1 and SVD2 samples. Table~\ref{tab:wrongtag} lists the fit result
for the wrong tag fractions in the SVD2 sample.  Those from the SVD1
sample have very similar values.
\begin{table}[htb]
\caption{
  Fit results for the wrong tag fractions of the SVD2 sample. They are
 determined in each $r$ bin as the averages and differences of the two
 flavours. 
}
\label{tab:wrongtag}
\begin{tabular}{ccc}
\hline \hline
$r$ & $(w_+ + w_-)/2$ & $(w_+ - w_-)$ \\
\hline
 0.000~--~0.250~~~~ & 0.461$\pm$0.006~ &~ $+0.001 \pm 0.009$ \\
 0.250~--~0.500~~~~ & 0.327$\pm$0.009~ &~ $-0.026 \pm 0.013$ \\
 0.500~--~0.625~~~~ & 0.216$\pm$0.010~ &~ $+0.032 \pm 0.015$ \\
 0.625~--~0.750~~~~ & 0.163$\pm$0.009~ &~ $-0.001 \pm 0.014$ \\
 0.750~--~0.875~~~~ & 0.124$\pm$0.009~ &~ $-0.020 \pm 0.014$ \\
 0.875~--~1.000~~~~ & 0.032$\pm$0.005~ &~ $+0.012 \pm 0.009$ \\
\hline \hline
\end{tabular}
\end{table}

We then perform fits to determine the values of $S^\pm$   
by fixing $\tau_{B^0}$ and  $\Delta m_d$  to the world 
average values and using previously determined $w_\pm$ and $S^\pm_\mathrm{tag}$
in each $r$ bin. The results are 
$S^+ (D^* \pi) = 0.050 \pm 0.029$, 
$S^- (D^* \pi) = 0.028 \pm 0.028$, 
$S^+ (D \pi)   = 0.031 \pm 0.030$, and 
$S^- (D \pi)   = 0.068 \pm 0.029$. 
The errors are statistical only.

\begin{figure}[!thb]
  \begin{center}
    \begin{tabular}{cc}
      \begin{minipage}{4.25cm}
        \begin{center} 
          \includegraphics[width=4.25cm,clip]{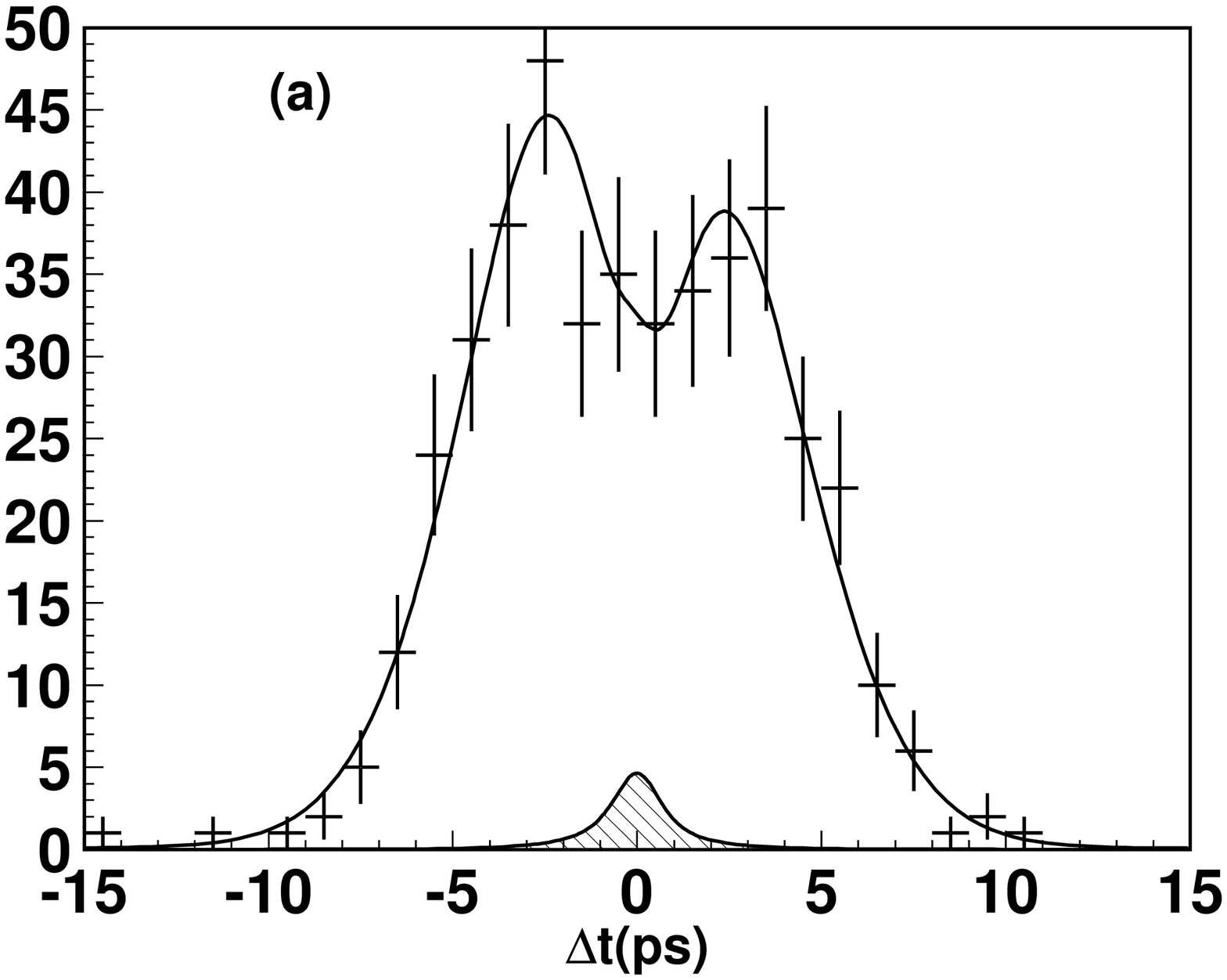} 
        \end{center}
      \end{minipage}
      &
      \begin{minipage}{4.25cm}
        \begin{center} 
          \includegraphics[width=4.25cm,clip]{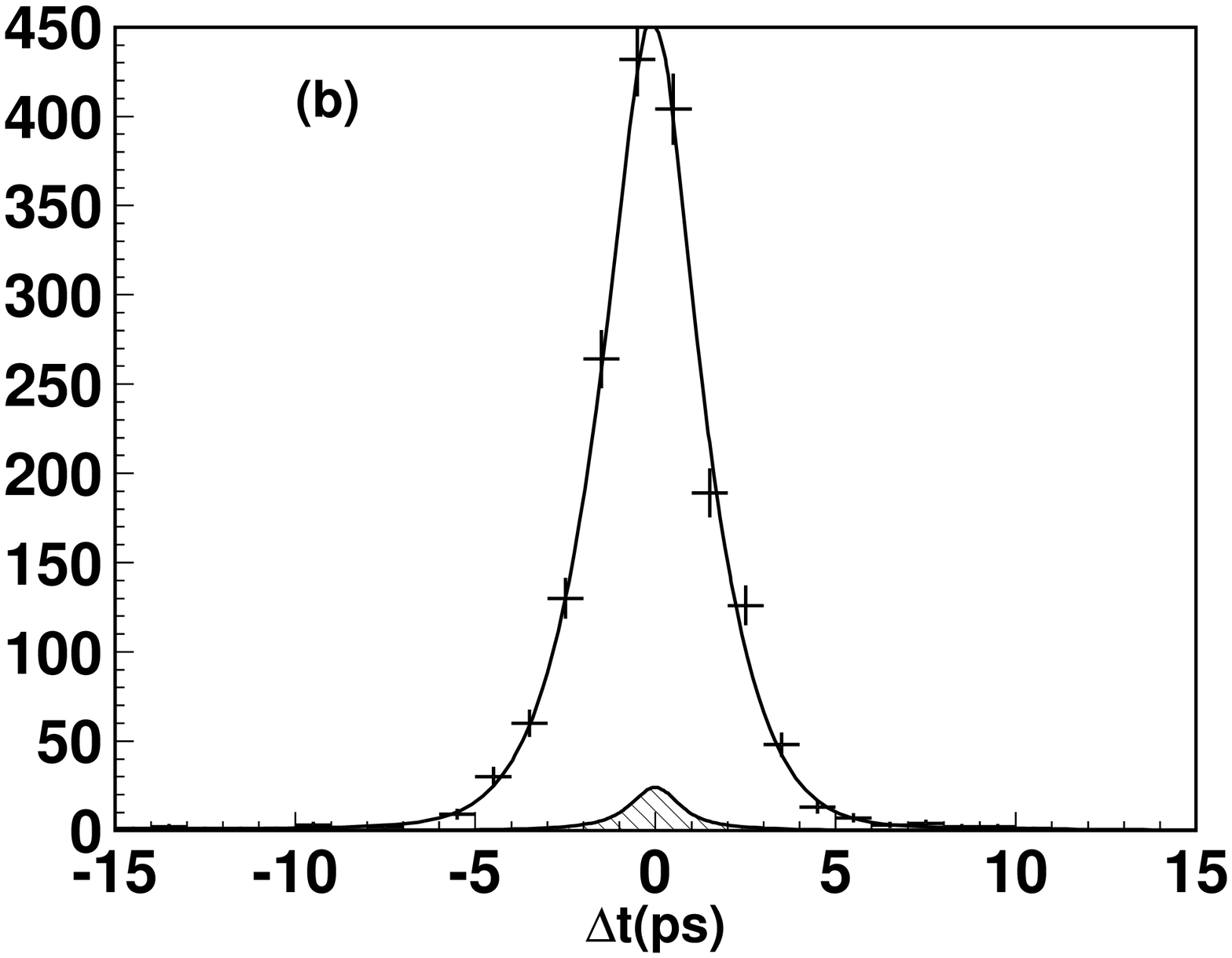} 
        \end{center}
      \end{minipage}
      \\ 
      \begin{minipage}{4.25cm}
        \begin{center} 
           \includegraphics[width=4.25cm,clip]{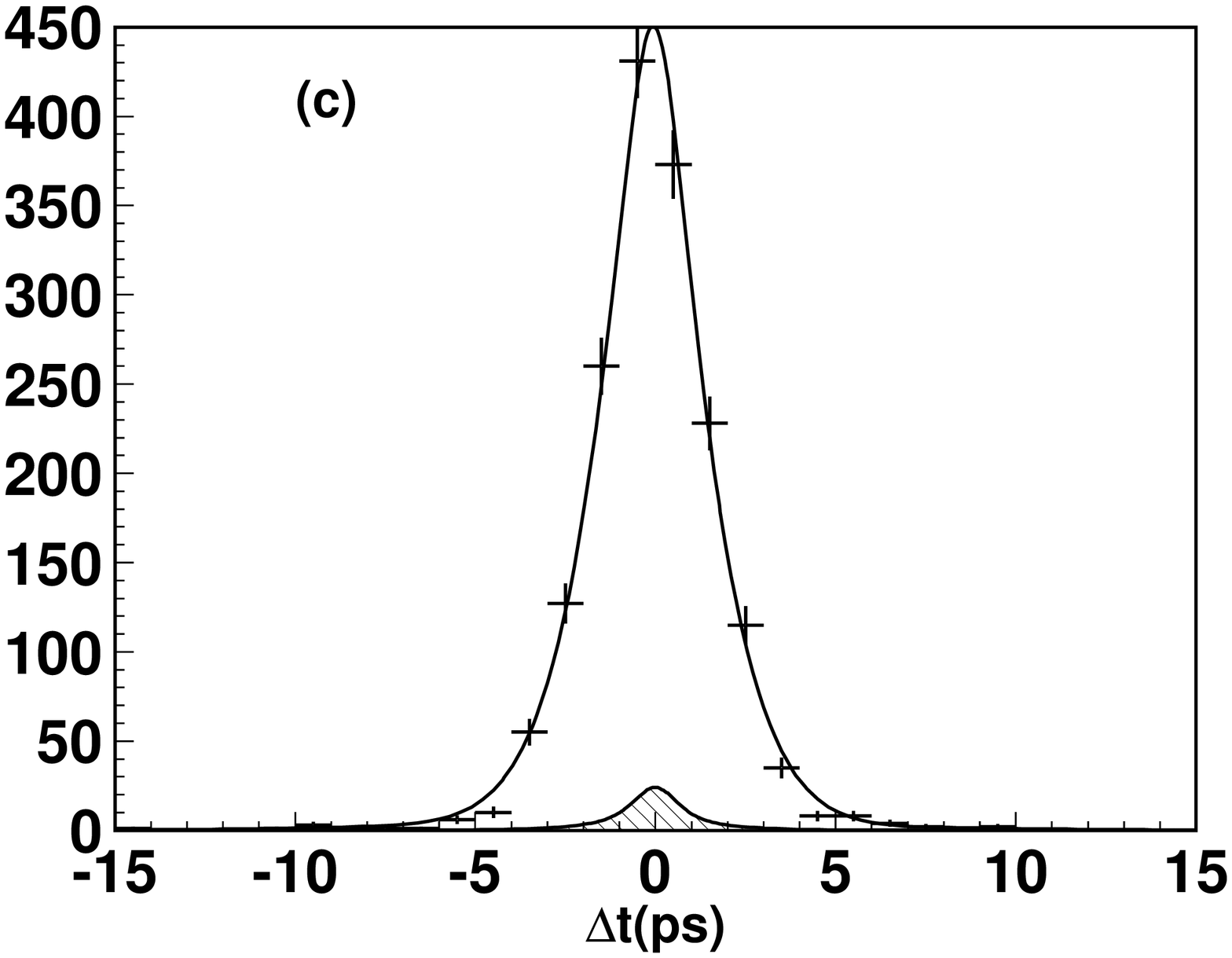} 
        \end{center}
      \end{minipage}
      &
      \begin{minipage}{4.25cm}
        \begin{center} 
      \includegraphics[width=4.25cm,clip]{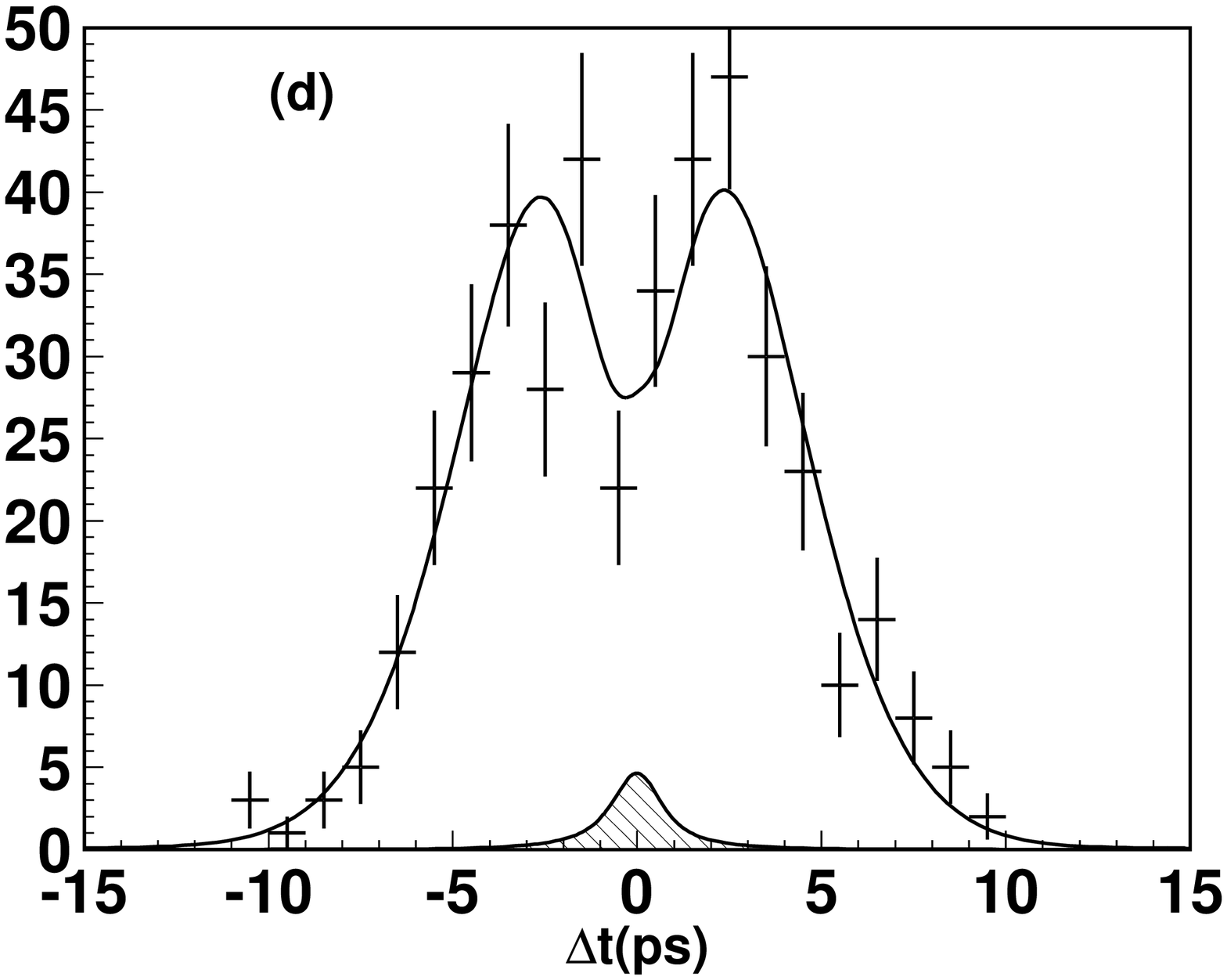} 
        \end{center}
      \end{minipage}
    \end{tabular}
  \end{center}
\vspace*{-5mm}  
\caption{
    $\Delta t$ distributions for the $D^* \pi$ events in the 
    $0.875 < r \leq 1.0$ flavour tagging quality bin.
    (a) $B^0 \to D^{*+} \pi^-$, 
    (b) $B^0 \to D^{*-} \pi^+$, 
    (c) $\overline{B}{}^0 \to D^{*+} \pi^-$, 
    (d) $\overline{B}{}^0 \to D^{*-} \pi^+$.
    Curves show the results of fits to the entire event sample, 
    and hatched regions indicate the backgrounds.
  }
    \label{dt_dstpi_r6}
  \begin{center}
    \begin{tabular}{cc}
       \begin{minipage}{4.25cm}
        \begin{center} 
          \includegraphics[width=4.25cm,clip]{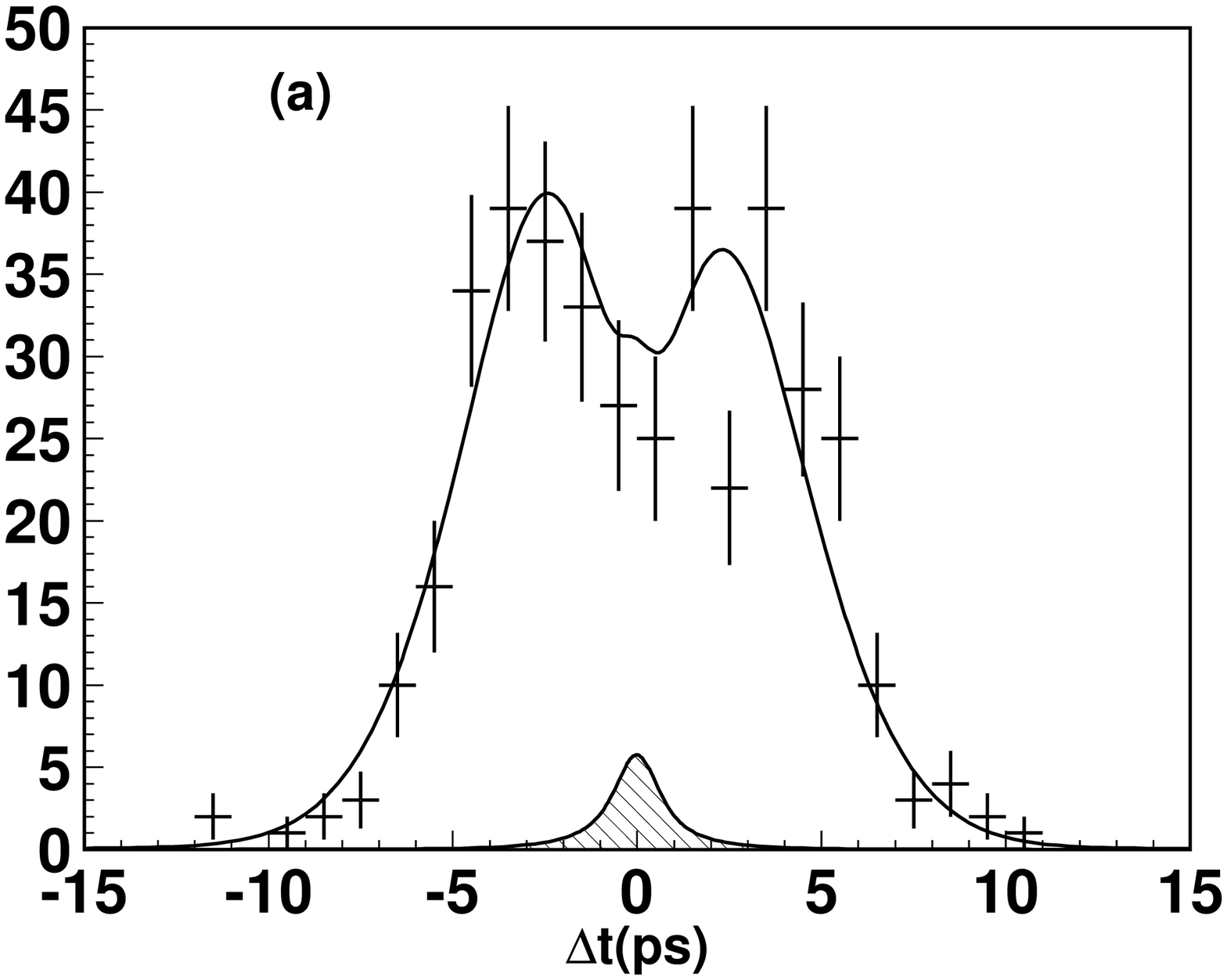} 
        \end{center}
      \end{minipage}
      &
      \begin{minipage}{4.25cm}
        \begin{center} 
           \includegraphics[width=4.25cm,clip]{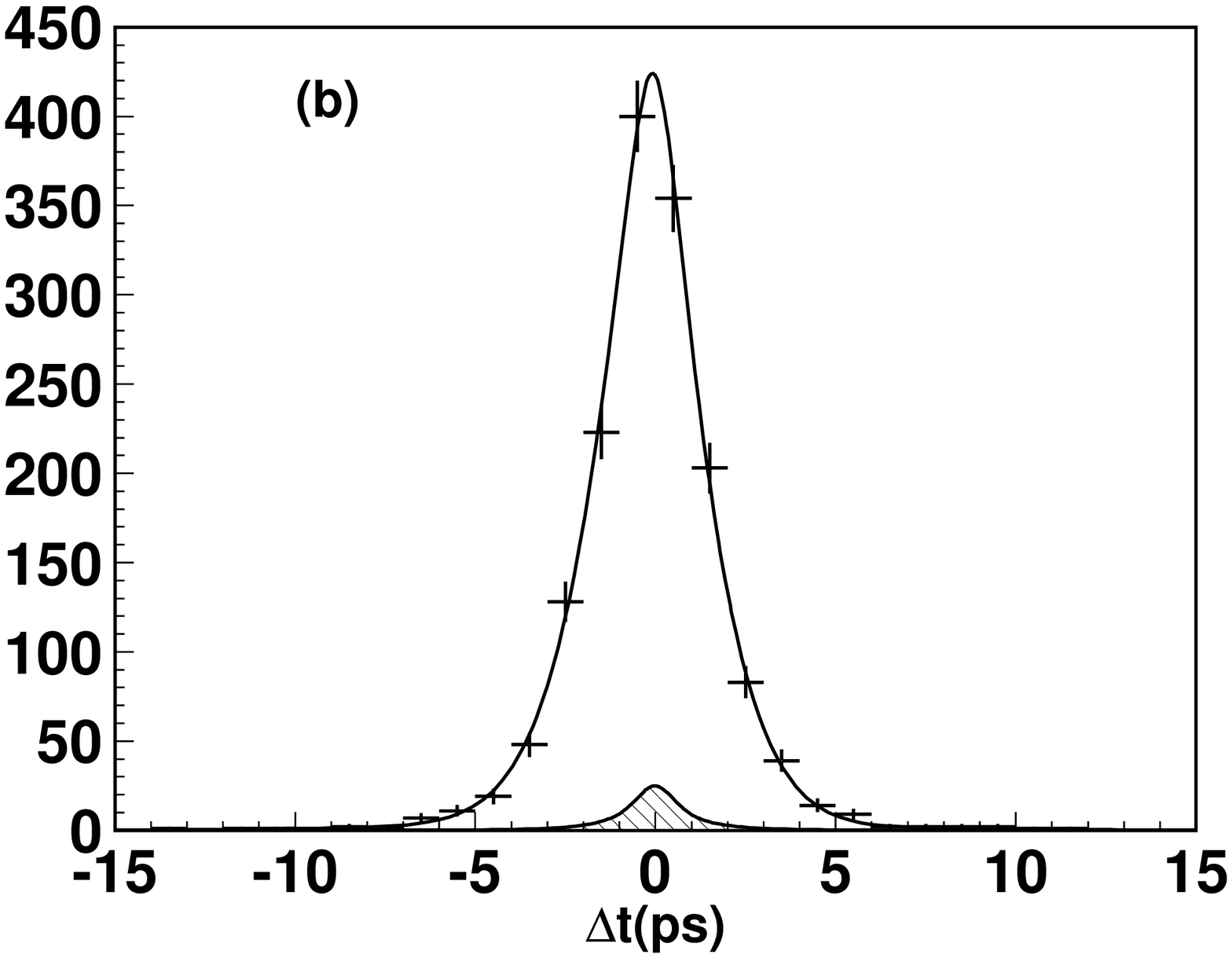} 
        \end{center}
      \end{minipage}
      \\
      \begin{minipage}{4.25cm}
        \begin{center} 
          \includegraphics[width=4.25cm,clip]{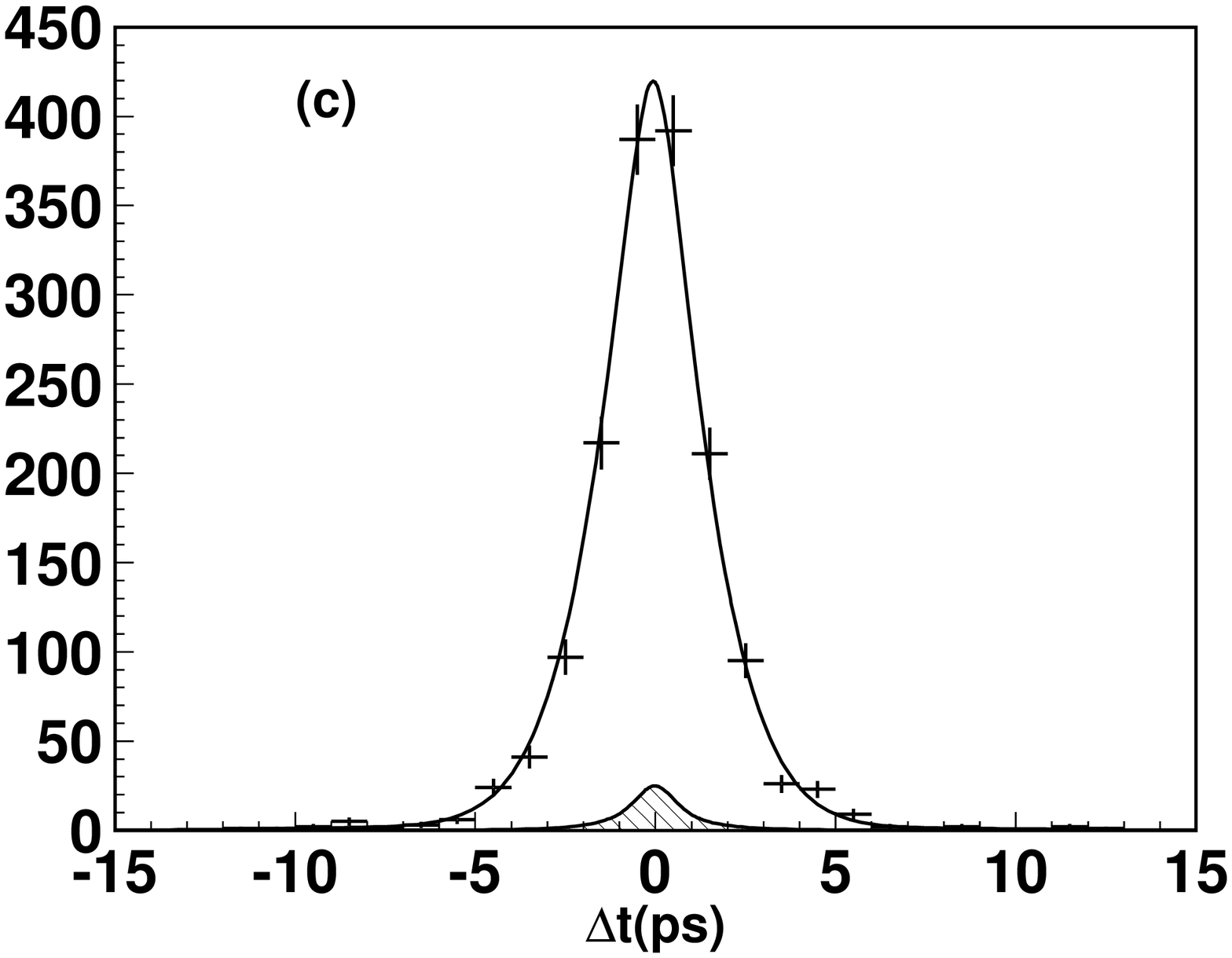} 
        \end{center}
      \end{minipage}
      &
      \begin{minipage}{4.25cm}
        \begin{center} 
           \includegraphics[width=4.25cm,clip]{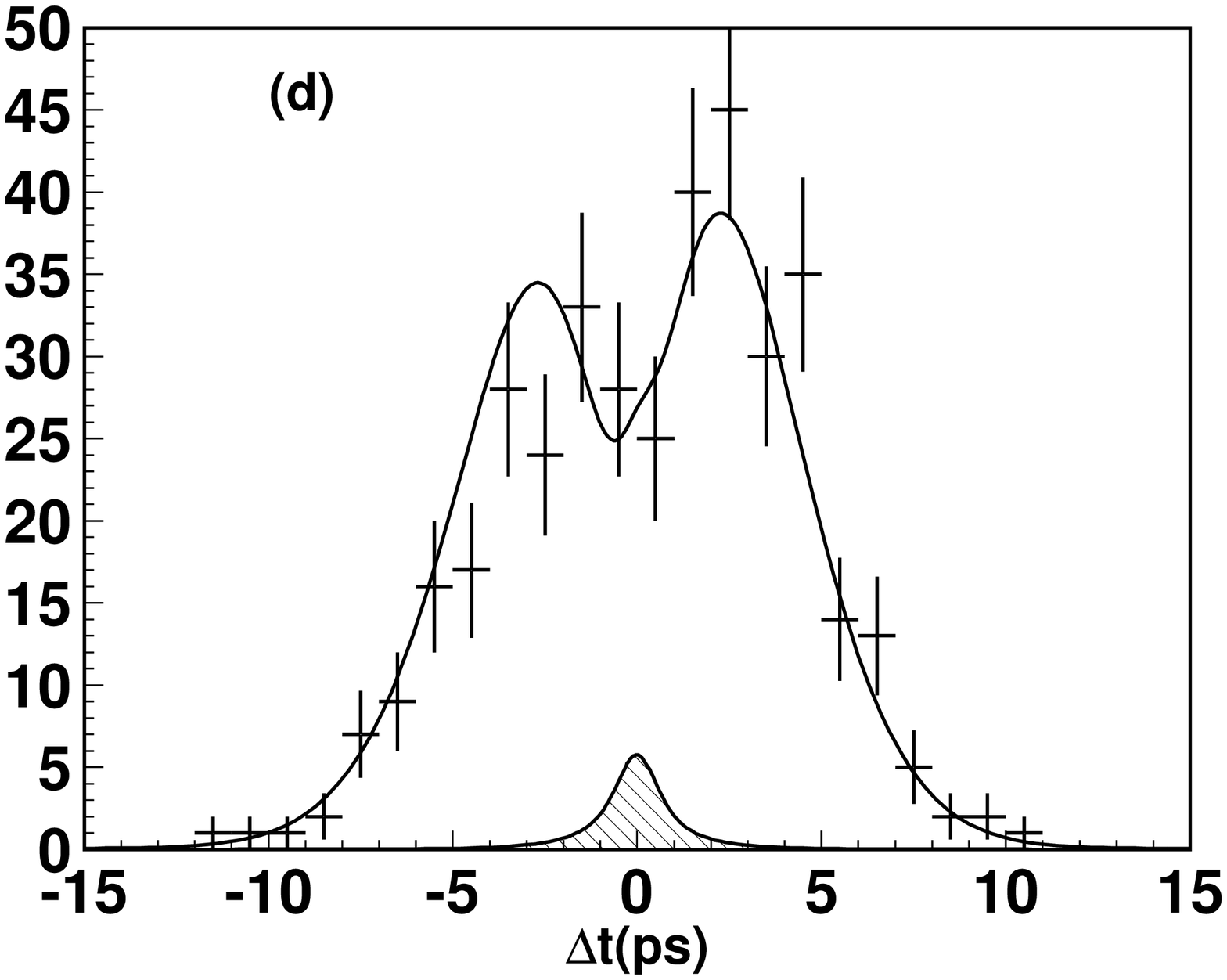} 
        \end{center}
      \end{minipage}
    \end{tabular}
  \end{center}
  \vspace*{-5mm}
  \caption{
     $\Delta t$ distributions for the $D \pi$ events in the 
    $0.875 < r \leq 1.0$ flavour tagging quality bin.
    (a) $B^0 \to D^{+} \pi^-$, 
    (b) $B^0 \to D^{-} \pi^+$, 
    (c) $\overline{B}{}^0 \to D^{+} \pi^-$, 
    (d) $\overline{B}{}^0 \to D^{-} \pi^+$.
    Curves show the results of fits to the entire event sample, 
   and hatched regions indicate the backgrounds.
  }
   \label{dt_dpi_r6}
\end{figure}

The $\Delta t$ distributions for the subsamples having the best quality 
flavour tagging ($0.875 < r \leq 1.0$) are shown in 
Fig.~\ref{dt_dstpi_r6} for the $D^* \pi$ mode and in
Fig.~\ref{dt_dpi_r6} for the $D \pi$ mode. 
For both modes, this region constitutes roughly 14\% of the total, but 
has significant statistical power for the $S^\pm$ 
determination. 

\subsection{Systematic Error}
The systematic errors come from the uncertainties of parameters 
that are constrained in the fit ($\Delta t$ resolution, background 
$\Delta t$ shape, background fractions, wrong tag fractions, 
vertexing, physics parameters), uncertainties due to the tag side asymmetry 
and biases induced by the fitting method.
To estimate contributions from the fit parameter uncertainties,          
we repeat the fit by varying each parameter by a given amount,             
and assign the shift in the $S^\pm$ parameters from the nominal fit as 
a systematic error.  
The signal $\Delta t$ resolution parameters 
are varied by $\pm 1\,\sigma$ of their errors.
       The background $\Delta t$ shape parameters are varied by 
$\pm 1\,\sigma$ of their errors. 
     For the background fraction,  in addition to 
varying the parameters of the $\Delta E$-$M_\mathrm{bc}$ signal region
fit by $\pm 1\,\sigma$, 
we vary the $\Delta E$ signal region cut by $\pm 5\,\mathrm{MeV}$
and $M_\mathrm{bc}$ signal region cut by $\pm 3\,\mathrm{ MeV}/c^2$ and 
the quadratic sum of these is assigned as an error.
Contributions from the peaking background are found to be negligible
based on a MC study.
     We vary the wrong tag fraction parameters by $\pm 1\,\sigma$ 
in each $r$ bin and add in quadrature.
 We vary the cut of $\xi$ by 
$+100$ and $-50$, which gives 0.004 for $D^*\pi$ and 0.002
for $D\pi$ as errors due to the vertexing. 
We repeat the fit after removing  
the $|\Delta t| < 70\,\mathrm{ps}$ cut and find a negligible shift.
  We vary the errors for $\tau_{B^0}$ and $\Delta m$ by
$\pm 1\,\sigma$.

For the tag side asymmetry, the quadratic
sum of statistical and systematic errors 
of $S^\pm_{\mathrm tag}$ parameters are varied by $\pm 1\,\sigma$ in each $r$
bin and the deviations are added in quadrature, giving 0.005 as an error.
Since this error includes contributions from unknown effects in the
vertex measurement such as the misalignment of the SVD and drift of the IP
profile, we do not explicitly assign additional error to the vertexing
other than that from the $\xi$ cut.  

Fit bias is tested using a $D^*\pi$ signal MC sample where one B meson
decays to $D^* \pi$, and the other B decays generically. We fit the
$S^\pm$ parameter to this sample, with and without the 
$S^\pm_\mathrm{tag}$ correction	given in  Eq.~\ref{eqntag}, taking 
$S^\pm_\mathrm{tag}$ values from a $B \to D^* l\nu$ signal MC sample. 
Neither MC sample includes the tag side $CP$ violation effects,
so the two fits should return the same $S^\pm$ values in principle. We
take the difference between the fits, 0.010, as a measure of
biases in the fitting procedure.

We obtain a total systematic error of 0.013 for $D^*\pi$
and 0.012 for $D\pi$.
Table~\ref{syserror} summarizes the contributions to the systematic errors. 
\begin{table}[htb]
\caption{
  Summary of the systematic errors in the $S^\pm$ measurements using the 
  full reconstruction method.
}
\label{syserror}
\begin{tabular}{lcc}
\hline \hline
Sources   
 &  $D^* \pi$  &  $D \pi$ \\
\hline
Signal $\Delta t$ resolution       & 0.005    & 0.004   \\
Background $\Delta t$ shape        & negligible~    & negligible   \\
Background fraction                & 0.002    & 0.001   \\
Wrong tag fraction                 & 0.002    & 0.002   \\
Vertexing                          & 0.004    & 0.002   \\
Physics parameters ($\Delta m, \tau_{B^0}$)    
                                   & 0.001    & 0.001   \\
Tag side asymmetry             & 0.005    & 0.005   \\
Fit bias                    & 0.010    & 0.010    \\
\hline
Total                           & 0.013    & 0.012   \\
\hline \hline
\end{tabular}
\end{table}

\subsection{Result}
We obtain
\begin{eqnarray}
S^+ (D^* \pi) &=& 0.050 \pm 0.029 \pm 0.013, \nonumber \\
S^- (D^* \pi) &=& 0.028 \pm 0.028 \pm 0.013, \nonumber \\
S^+ (D \pi) &=& 0.031 \pm 0.030 \pm 0.012, \nonumber \\
S^- (D \pi) &=& 0.068 \pm 0.029 \pm 0.012 
\end{eqnarray}
from the full reconstruction method, where the first error is 
statistical and the second error is systematic. 

\section{Partial Reconstruction Analysis}
\subsection{Signal Event Selection}
Candidate events are selected by requiring the presence of
fast pion and slow pion candidates.
In order to obtain accurate vertex position determinations,
fast pion candidates are 
required to have a radial (longitudinal) impact parameter $dr <
0.1\,\mathrm{cm}$ ($|dz| < 2.0\,\mathrm{cm}$), 
to have associated hits in the SVD,
and to have a polar angle in the laboratory frame in the range
$30^\circ < \theta_{\rm lab} < 135^\circ$.
The vertex positions are obtained by fits 
of the candidate tracks with the IP. 
Fast pion candidates are required to be inconsistent 
with the lepton hypothesis (see below), 
and the kaon hypothesis,
based on information from the CDC, TOF and ACC. 
A requirement on the fast pion cms momentum of
$1.83 \, {\rm GeV}/c < p_{\pi_f} < 2.43 \, {\rm GeV}/c$ is made;
this range includes both signal and sideband regions (defined below).
Slow pion candidates are required to have cms momentum in the range
$0.05 \, {\rm GeV}/c < p_{\pi_s} < 0.30 \, {\rm GeV}/c$.
No requirement is made on particle identification for slow pions; 
since they are not used for vertexing, only a loose requirement 
that they originate from the IP is made.
The fast and slow pion candidates must have opposite charges.

\subsection{Flavour Tagging}
In order to tag the flavour of the associated $B$ meson and to reduce 
background from continuum 
$e^+e^- \to q\overline{q} \ (q = u,d,s,c)$ processes,
we require the presence of a high-momentum lepton in the event.
Tagging lepton candidates are required to be positively identified
either as electrons, on the basis of information from the CDC, ECL and ACC, 
or as muons, on the basis of information from the CDC and the KLM.
They are required to have momentum in the range
$1.2 \ {\rm GeV}/c < p_{l_{\rm tag}} < 2.3 \ {\rm GeV}/c$,
and to have an angle with the fast pion candidate that satisfies
$-0.75 < \cos \delta_{\pi_f l}$ in the cms.
The lower bound on the momentum and the requirement on the angle 
also reduce, to a negligible level,
the contribution of leptons produced from semi-leptonic
decays of the unreconstructed $D$ mesons in the $B^0 \to D^{*\mp}\pi^\pm$ decay chain.
No other tagging lepton candidate with momentum greater than 
$1.0 \ {\rm GeV}/c$ is allowed in the event to reduce the 
mistagging probability, and also to reduce the contribution
from leptonic charmonium decays.
Identical vertexing requirements to those for fast pion candidates 
are made in order to obtain an accurate $z_{\rm tag}$ position.
To further suppress the small remaining continuum background,
we impose a loose requirement on the ratio of 
the second to zeroth Fox-Wolfram~\cite{fw} moments, $R_2 < 0.6$.

\subsection{Kinematic Fit}
Signal events are distinguished from background using three kinematic
variables, which are approximately independent for signal.
These are denoted by $p_{\pi_f}$, $\cos \delta_{\pi_f \pi_s}$ and 
$\cos \theta_{\rm{hel}}$.
For signal, the fast pion cms momentum, $p_{\pi_f}$,
has a uniform distribution, smeared by the experimental resolution,
as the fast pion is monoenergetic in the $B$ rest frame.
The cosine of the angle between the fast pion direction and the 
opposite of the slow pion direction in the cms, $\cos \delta_{\pi_f \pi_s}$,
peaks sharply at $+1$ for signal,
as the slow pion follows the $D^*$ direction  
due to the small energy released in the $D^*$ decay.
The cosine of the angle between the slow pion direction and the opposite
of the 
$B$ direction in the $D^*$ rest frame, $\cos \theta_{\rm{hel}}$,
has a distribution proportional to $\cos \theta_{\rm{hel}}^2$ for signal 
events,
as the $B$ decay is a pseudoscalar to pseudoscalar vector transition.
Since the $D^*$ is not fully reconstructed, 
$\cos \theta_{\rm{hel}}$ is calculated using kinematic constraints, 
and the background can populate the unphysical region 
$\left|\cos \theta_{\rm{hel}}\right|>1$.

We select candidates that satisfy
$1.93 \ {\rm GeV}/c < p_{\pi_f} < 2.43 \ {\rm GeV}/c$, 
$0.850 < \cos \delta_{\pi_f \pi_s} < 1.000$ and 
$-1.70 < \cos \theta_{\rm{hel}}  < 1.80$.
In the cases where more than one candidate satisfies these criteria,
we select the one with the largest value of $\cos \delta_{\pi_f \pi_s}$.
We further define signal regions in $p_{\pi_f}$ and $\cos \delta_{\pi_f \pi_s}$ as 
$2.13 \ {\rm GeV}/c < p_{\pi_f} < 2.43 \ {\rm GeV}/c$, 
$0.925 < \cos \delta_{\pi_f \pi_s} < 1.000$,
and two regions in $\cos \theta_{\rm{hel}}$:
$-1.00 < \cos \theta_{\rm{hel}}  < -0.30$ and 
$+0.40 < \cos \theta_{\rm{hel}}  < +1.10$.

Background events are separated into three categories:
$D^{*\mp}\rho^{\pm}$, which is kinematically similar to the signal; 
correlated background, in which the slow pion originates from the decay of 
a $D^*$ that in turn originates from the decay of the same $B$
candidate as the fast pion candidate ({\it e.g.}, $D^{**}\pi$);
and uncorrelated background, which includes everything else
({\it e.g.}, continuum processes, $D\pi$).
The kinematic distributions of the background categories and the signal
are determined from a large MC sample,
corresponding to two times the integrated luminosity of our data sample,
in which the branching fractions of the signal and major background 
sources are weighted according to the 
most recent knowledge~\cite{PDG,kuzmin}.
We also use this MC sample for various tests of the analysis algorithms.

\begin{figure}[!htb]
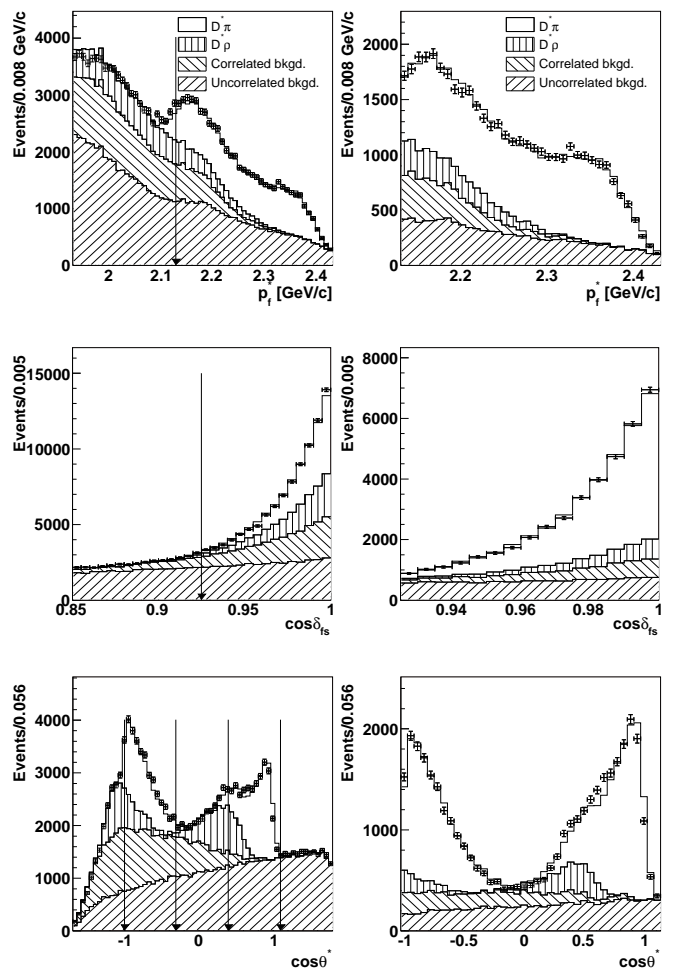

  \includegraphics[width=4.25cm,clip]{fig1a.epsi}   
  \includegraphics[width=4.25cm,clip]{fig1d.epsi}

\vspace*{0.5cm}
  \includegraphics[width=4.25cm]{fig1b.epsi}   
  \includegraphics[width=4.25cm]{fig1e.epsi}

\vspace*{0.5cm}
  \includegraphics[width=4.25cm]{fig1c.epsi}
  \includegraphics[width=4.25cm]{fig1f.epsi}

  \caption{
    \label{fig:kin_fit}
    Results of the kinematic fits to $D^*\pi$ candidates,
    projected onto (top) the $p_{\pi_f}$
    axis, (middle) the $\cos \delta_{\pi_f \pi_s}$ axis and the
    (bottom) $\cos \theta_{\rm{hel}}$ axis, in both (left) selection
    region and (right) signal region.
    The arrows show the edges of the signal regions.
  }
\end{figure}

Event-by-event signal fractions are determined from 
binned maximum likelihood fits to the three-dimensional 
kinematic  distributions
(6 bins of $p_{\pi_f}$ $\times$ 5 bins of $\cos \delta_{\pi_f \pi_s}$ 
$\times$ 10 bins of $\cos \theta_{\rm{hel}}$). 
The results of these fits, projected onto each of the three variables,
are shown in Fig.~\ref{fig:kin_fit},
and summarized in Table~\ref{tab:kin_fit}.

\begin{table}[htb]
   \caption{
     \label{tab:kin_fit}
     Summary of the results of the three-dimensional fits to
     kinematic variables.
     The numbers of events and fractions given for each category
     are those extrapolated to inside
     the signal regions in all three variables. The errors on the  
fractions
     are returned by the fit and propagated to the corresponding  
numbers of
     events.
   }
   \begin{center}
     \begin{tabular}{lcc}
\hline \hline
              &  Candidates &   Fraction \\
\hline \hline
Data         & $ 32844\pm181 $ & $-$ \\
$D^* \pi$    & $ 21741\pm217 $ & $0.662\pm0.007$ \\
$D^* \rho$   & $  2091\pm 45 $ & $0.064\pm0.001$ \\
Correlated background   & $  2703\pm 42 $ & $0.082\pm0.001$ \\
Uncorrelated background    & $  6287\pm 40 $ & $0.191\pm0.001$ \\
\hline \hline
     \end{tabular}
   \end{center}
\end{table}

\subsection{\boldmath $\Delta t$ Fit Procedure}
In order to measure the $CP$ violation parameters in the $D^*\pi$ sample,
we perform a simultaneous unbinned fit to the 
same-flavor (SF) events, in which the fast pion and the tagging lepton
have the same charge, and opposite-flavor (OF) events, in which the
fast pion and the tagging lepton have the opposite charge. 
We minimize the quantity $-2\ln {\cal L} = -2 \sum_i \ln {\cal L}_i$,
where 
\begin{equation}
  \label{eq:likelihood}
  {\cal L}_i = 
  f_{D^*\pi} P_{D^*\pi} + f_{D^*\rho} P_{D^*\rho} +
  f_{\rm unco} P_{\rm unco} + f_{\rm corr} P_{\rm corr}.
\end{equation}

The event-by-event signal and background fractions
(the $f$ terms) are taken from the results of the kinematic fits.
Each $P$ term contains an underlying physics PDF,
with experimental effects taken into account. 
For $D^*\pi$ and $D^*\rho$, the PDF is given by Eq.~\ref{eq:evol},
where for $D^*\rho$ the terms $S^\pm$ are effective parameters
averaged over the helicity states~\cite{dstarrho}
that are constrained to be zero.
The PDF for correlated background contains
a term for neutral $B$ decays 
(given by Eq.~\ref{eq:evol} with $S^\pm = 0$),
and, in the case of OF events, a term for charged $B$ decays
(for which the PDF is 
$\frac{1}{2\tau_{B^+}} e^{-\left| \Delta t \right| / \tau_{B^+}}$,
where $\tau_{B^+}$ is the lifetime of the charged $B$ meson).

The PDF for uncorrelated background also contains
neutral and charged $B$ components, with the remainder from continuum 
$e^+e^- \to q\overline{q} \ (q = u,d,s,c)$ processes.
The continuum PDF is modelled with two components: 
one with negligible lifetime,  and the other with a finite lifetime.
The sideband parameters are determined from data sidebands, 
as described later.

As mentioned above, experimental effects need to be taken into
account to obtain the $P$ terms of Eq.~\ref{eq:likelihood}.
Mistagging is taken into account using
\begin{eqnarray}
    \label{eq:exp_pdf}
 \lefteqn{ P( l_\mathrm{tag}^-, \pi_f^\pm)  = } \nonumber \\  
  && ( 1 - w_- ) P(B^{0} \to D^{*\mp} \pi^\pm)       
  +  w_+ P(\overline{B}{}^0 \to D^{*\mp} \pi^\pm) \nonumber \\
 \lefteqn{ P( l_\mathrm{tag}^+, \pi_f^\pm)  = } \nonumber \\  
  && ( 1 - w_+ ) P(\overline{B}{}^0 \to D^{*\mp} \pi^\pm) 
  +  w_- P(B^{0} \to D^{*\mp} \pi^\pm),\nonumber \\
  \end{eqnarray}
where $w_+$ and $w_-$ are the wrong tag fractions,
and are determined from the data as free parameters in the fit for $S^\pm$.
It should be noted that the $w_+$ and $w_-$ values used here are 
different from those used in the full reconstruction methods because the
flavour-tagging methods differ in two cases. 

The time difference $\Delta t$ is related to the measured quantity $\Delta z$
as described in Eq.~\ref{eq:dtvsdz}, with an additional term due to 
possible offsets in the mean value of $\Delta z$,
\begin{equation}
  \label{eq:dt_offset}
  \Delta t \longrightarrow \Delta t + \epsilon_{\Delta t} \simeq \left( \Delta z + \epsilon_{\Delta z} \right) / \beta\gamma c.
\end{equation}
It is essential to allow non-zero values of $\epsilon$ since a 
small bias can mimic the effect of $CP$ violation:
\begin{equation}
      \cos (\Delta m \Delta t) 
 \to
      \cos (\Delta m \Delta t) - 
\Delta m \epsilon_{\Delta t} \sin (\Delta m \Delta t).
\end{equation}
A bias as small as $\epsilon_{\Delta z} \sim 1 \ \mu{\rm m}$ can lead to 
sine-like terms as large as $0.01$, 
comparable to the expected size of the $CP$ violation effect.
We allow separate offsets for each combination of fast pion and 
tagging lepton charges.
We also apply a small correction to each measured vertex position
to correct for a known bias due to the relative misalignment of the SVD 
and CDC in
SVD1 data. This correction is dependent on the track charge, 
momentum and polar angle, measured in the laboratory frame.
It is obtained by comparing the vertex positions 
calculated with the alignment constants used in the data,
to those obtained with an improved set of alignment constants~\cite{dzb}.
The alignment in SVD2 data was found to be comparable to that of
the corrected SVD1 data. No additional correction was thus applied to SVD2 data.

\subsection{\boldmath $\Delta t$ Resolution}
Resolution effects are taken into account in a way similar to our
full reconstruction analysis.
The algorithm includes components related to detector resolution,
kinematic smearing and non-primary tracks.

For correctly tagged signal events, both the fast pion and 
the tagging lepton originate directly from $B$ meson decays. 
Therefore we do not include any additional smearing due to non-primary 
tracks in
these events. Incorrectly tagged events, however, almost exclusively
originate from secondary leptons. Furthermore, due 
to the kinematic constraints on the momentum of the fast pion, secondary 
tracks consist almost exclusively of wrong-tag leptons. Only uncorrelated
background events contain a small amount of secondary pions, which also
give the wrong flavour information. In order to take
this effect into account, the PDFs of incorrectly tagged events are
convolved with an additional resolution component whose parameters
are determined from MC simulations. Three different sets 
of parameters are used: one set for uncorrelated
background in the uncorrelated background sideband, where the fast pion
momentum cut is loosened; one set for uncorrelated background in the two other regions (signal region and correlated background side-band); one set for all
other categories of events (signal, $D^{*}\rho$ and correlated background),
which contain similar amounts of secondary leptons, and no secondary pions. Because of the aforementioned correlation between mistagging and non-primary
tracks, each of these sets are linked to different wrong-tag fractions 
$w_{\pm}$. 

The effect of the approximation that the $B$ mesons are at rest
in the cms in Eq.~\ref{eq:dtvsdz} is taken into account~\cite{vertex_nim}.
We use a slightly modified algorithm to describe the detector resolution,
in order to precisely describe the observed behaviour for 
single track vertices. The resolution for each track is described by the sum 
of three Gaussian components, with a common mean of zero,
and widths that are given by the measured vertex error for each track
multiplied by different scale factors.

We measure the five parameters of the detector resolution function 
(three scale factors and two parameters giving the relative normalizations
of the Gaussians) using $J/\psi \to \mu^+\mu^-$ candidates.
These are selected using criteria similar to those for $D^*\pi$,
except that both tracks are required to be identified as muons, 
and their invariant mass is required to be consistent with that of 
the $J/\psi$.
Vertex positions are obtained independently for each track,
in the same way as described above.
The quantity $\Delta z = z_{\mu^+} - z_{\mu^-}$ then describes the
detector resolution,
which is the convolution of the two vertex resolutions,
since for $J/\psi \to \mu^+\mu^-$ the underlying PDF is a delta function.

\begin{figure}[!htb]
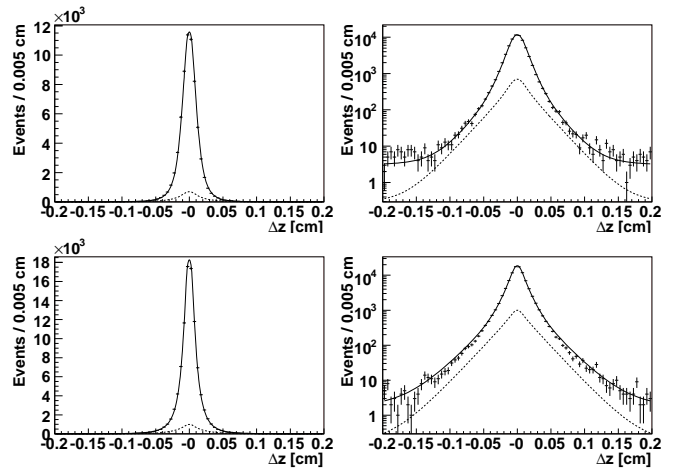

  \begin{center}
    \includegraphics[width=4.25cm]{fig2a.epsi}   
    \includegraphics[width=4.25cm]{fig2c.epsi}\\  
    \includegraphics[width=4.25cm]{fig2b.epsi}   
    \includegraphics[width=4.25cm]{fig2d.epsi}\\  
  \end{center}
  \caption{
    \label{fig:fit_jpsi}
    Result of the resolution parameter extraction procedure
    for $J/\psi \to \mu^+\mu^-$ candidates selected from (top) SVD1
    and (bottom) SVD2 data,
    shown with both (left) linear and (right) logarithmic coordinate scales.
    The data points show the $\Delta z$ distribution for candidates
    in the signal region; the full line shows the result of the fit. The
 dashed line indicates the background. 
  }
\end{figure}

We perform an unbinned maximum likelihood fit using events 
in the $J/\psi$ signal region in the di-muon invariant mass,
and using sideband regions to determine the shape of the background 
under the peak.
The underlying $\Delta z$ PDF of the background is parametrized in the same way 
as that used for continuum, described above. Two sets of parameters are
simultaneously determined for SVD1 and SVD2 data.
We also take a possible offset into account, so that there are in total 16 free parameters 
in this fit (five describing the resolution function,
two describing the background, and one $\Delta z$ offset, each for SVD1 and SVD2 data).
The result is shown in Fig.~\ref{fig:fit_jpsi}.

\subsection{\boldmath Background $\Delta t$ Shape}
The free parameters in the background PDFs $P_{\rm unco}$ and $P_{\rm corr}$ 
are determined using data from sideband regions.
To measure the uncorrelated background shape,
we use events in a side-band region of
$-0.5 < \cos \delta_{\pi_f \pi_s} < 0.5$, which
by definition is only populated by uncorrelated background, thus
providing a very large sample to extract the various components of
this background. We perform a simultaneous fit to OF and SF candidates,
in both SVD1 and SVD2 data.

To obtain the correlated background parameters,
a simultaneous fit is carried out to OF and SF events 
in a sideband region of
$0.85 < \cos \delta_{\pi_f \pi_s} < 0.90$,
$1.93 \ {\rm GeV}/c < p_{\pi_f} < 2.13 \ {\rm GeV}/c$ and 
$-1.7 < \cos \theta_{\rm{hel}} < -0.3$.
This sideband region is dominated by correlated and 
uncorrelated backgrounds\textemdash{}the contributions 
from $D^*\pi$ and $D^*\rho$ are found to be small in MC.
The uncorrelated background parameters are fixed to the values
found in the previous fit, except the $\Delta z$ offset (which
are the same as that of correlated background) and the wrong-tag
fractions, as mentioned above.

\subsection{Fit Result}
In order to test our fit procedure,
we first constrain $S^+$ and $S^-$ to be zero and perform a fit in
which $\tau_{B^0}$ and $\Delta m$ 
(as well as two wrong tag fractions and eight offsets) are free parameters.
We obtain $\tau_{B^0} = 1.495 \pm 0.012 \ {\rm ps}$
and $\Delta m = 0.506 \pm 0.006 \ {\rm ps}^{-1}$,
where the errors are statistical only.
These values are compatible with the current world averages~\cite{PDG}.
Reasonable agreement with the input values is also obtained in MC.
Furthermore, fits to the MC with $S^\pm$ floated give results 
consistent with zero, as expected.

\begin{figure}[htb]
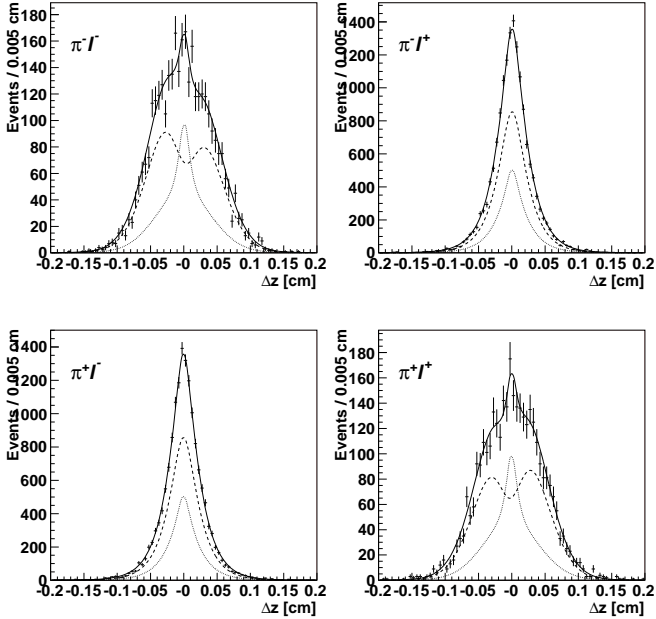

  \begin{center}
    \includegraphics[width=4.25cm]{fig3a.epsi}
    \includegraphics[width=4.25cm]{fig3b.epsi} 

\vspace*{0.5cm}
    \includegraphics[width=4.25cm]{fig3c.epsi}
    \includegraphics[width=4.25cm]{fig3d.epsi}

  \end{center}
  \caption{
    \label{fig:myfit}
    Results of the fit to obtain $S^+$ and $S^-$.
    The fit result is superimposed on the data.
    The signal component is shown as the dashed line.
     The dotted line indicates the background contribution.
  }
\end{figure}

\begin{figure}[htb]
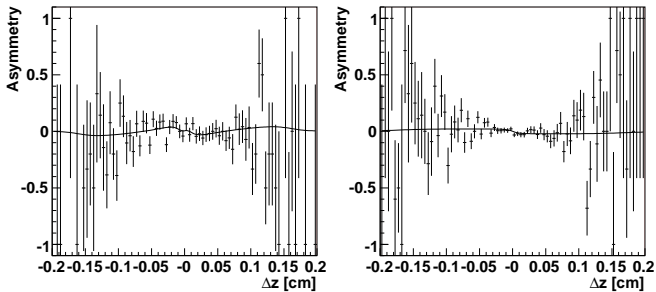

  \begin{center}
    \includegraphics[width=4.25cm]{fig5a.epsi}
    \includegraphics[width=4.25cm]{fig5b.epsi}
  \end{center}
  \caption{
    \label{fig:myfit_asp}
    Results of the fit to obtain $S^+$ and $S^-$,
    shown as asymmetries in the 
    (left) same flavour events and
    (right) opposite flavour events.
    The fit result is superimposed on the data.
  }
\end{figure}

To extract the $CP$ violation parameters we fix $\tau_{B^0}$ and 
$\Delta m$ at their world average values,
and fit with $S^+$, $S^-$, two wrong tag fractions and eight offsets
as free parameters.
We obtain
$S^+ = 0.048 \pm 0.028$ and $S^- = 0.034 \pm 0.027$
where the errors are statistical only.
The wrong tag fractions are $w_- = (4.8 \pm 1.6)\%$ and 
$w_+ = (3.4 \pm 1.6)\%$. 
All floating offsets are consistent with zero except  the one 
for the $\pi^{-}\ell^{+}$ combinations in the SVD1 sample.
The results are shown in Fig.~\ref{fig:myfit}.
To further illustrate the $CP$ violation effect,
we define asymmetries in the 
same flavour events (${\cal A}^{\mathrm{SF}}$) 
and in the opposite flavour events (${\cal A}^{\mathrm{OF}}$), as 
\begin{eqnarray}
  {\cal A}^{\mathrm{SF}} & = &
  \frac{ N_{\pi^-l^-}(\Delta z) - N_{\pi^+l^+}(\Delta z) } 
       { N_{\pi^-l^-}(\Delta z) + N_{\pi^+l^+}(\Delta z) }, \nonumber \\
  {\cal A}^{\mathrm{OF}} & = &
  \frac{ N_{\pi^+l^-}(\Delta z) - N_{\pi^-l^+}(\Delta z) } 
       { N_{\pi^+l^-}(\Delta z) + N_{\pi^-l^+}(\Delta z) }, 
\end{eqnarray}
where the $N$ values indicate the number of events 
for each combination of fast pion and tag lepton charge.
These are shown in Fig.~\ref{fig:myfit_asp}.
Note that due to the relative contributions of the
sine terms in Eq.~\ref{eq:evol}, 
vertex biases ({\it i.e.} non-zero offsets)
can induce an opposite flavour asymmetry,
whereas the same flavour asymmetry is more robust.

\subsection{Systematic Error}
Systematic errors due to the resolution function,
the background fractions, the background parameters, and physics
parameters are estimated by varying the values used in the fit by 
$\pm 1 \sigma$. 
Systematic effects due to differences between data and MC
in the distributions used in the kinematic fit are further investigated 
by repeating the fit using different binning.
We repeat the entire fit procedure using twice as many bins 
in each of the three discriminating variables.
Since $\cos \delta_{\pi_f \pi_s}$ is used in the best candidate selection,
we also repeat the algorithm without using this variable in the kinematic fit.
The largest deviation ($0.007$) is assigned as a systematic error.

The resolution function parameters are precisely determined 
from the fit to $J/\psi \to \mu^+\mu^-$ candidates.
We consider systematic effects due to our lack of knowledge
of the exact functional form of the resolution function:
using different parametrizations results in shifts of $S^\pm$ 
as large as $0.008$, which we assign as a systematic error.
Allowing for effective $S^\pm$ terms of $\pm0.05$ in the $D^*\rho$ and
correlated background PDFs leads to systematic errors of $0.003$ and
$0.002$, respectively. 

Vertex biases may lead to a large systematic error on $S^\pm$,
so we have introduced offsets to make our analysis relatively
insensitive to such biases. These additional free parameters cause an 
increase in the statistical error of about $20\%$.
In order to estimate the systematic error due to these offsets, we
compare the results in MC simulations with and without the offsets. A 
difference of $+0.011$ is found for $S^{+}$
and is assigned as a systematic error for the offsets.
We have further tested our fit routine for possible fit biases,
such as could be caused by neglecting terms of $R^2_{D^* \pi}$ in the fit,
by generating a number of large samples of signal MC 
with different input values of $S^+$ and $S^-$. All results come out
consistent with the input values, without evidence of any bias.

The systematic errors are summarized in Table~\ref{tab:systematics}.
The total systematic error ($0.017$) is obtained by adding the 
above terms in quadrature.

\begin{table}[htb]
  \caption{
    \label{tab:systematics}
    Summary of the systematic errors in the $S^\pm$ measurements using
 the partial reconstruction method.
  }
  \begin{center}
    \begin{tabular}{lc}
      \hline \hline
      Source & Error \\
      \hline
      Resolution fit                              & $0.002$ \\
      Resolution models                           & $0.008$ \\
      Kinematic smearing                          & $0.002$ \\      
      Non-primary tracks                          & $0.004$ \\
      Background shapes                           & negligible \\
      Kinematic fit                               & $0.007$ \\
      $\tau_{B^0}$, $\Delta m$                    & $0.001$ \\
      $CP$ violation in $D^*\rho$ and corr. bkgd. & $0.004$ \\
      Vertexing                                   & $0.011$ \\
      \hline
      Total                                       & $0.017$ \\
      \hline \hline
    \end{tabular}
  \end{center}
\end{table}

\subsection{Result}
The results using the partial reconstruction method are
\begin{eqnarray}
 S^+ & = & 0.048 \pm 0.028 \pm 0.017 ,  \nonumber \\
 S^- & = & 0.034 \pm 0.027 \pm 0.017 , 
\end{eqnarray}
where the first error is statistical and the second error is systematic.

\section{Discussion}
\subsection{\boldmath Final results of $S^\pm (D^* \pi)$ and  $S^\pm (D \pi)$}
\begin{figure}[htb]
  \begin{center}
    \includegraphics[width=0.235\textwidth]{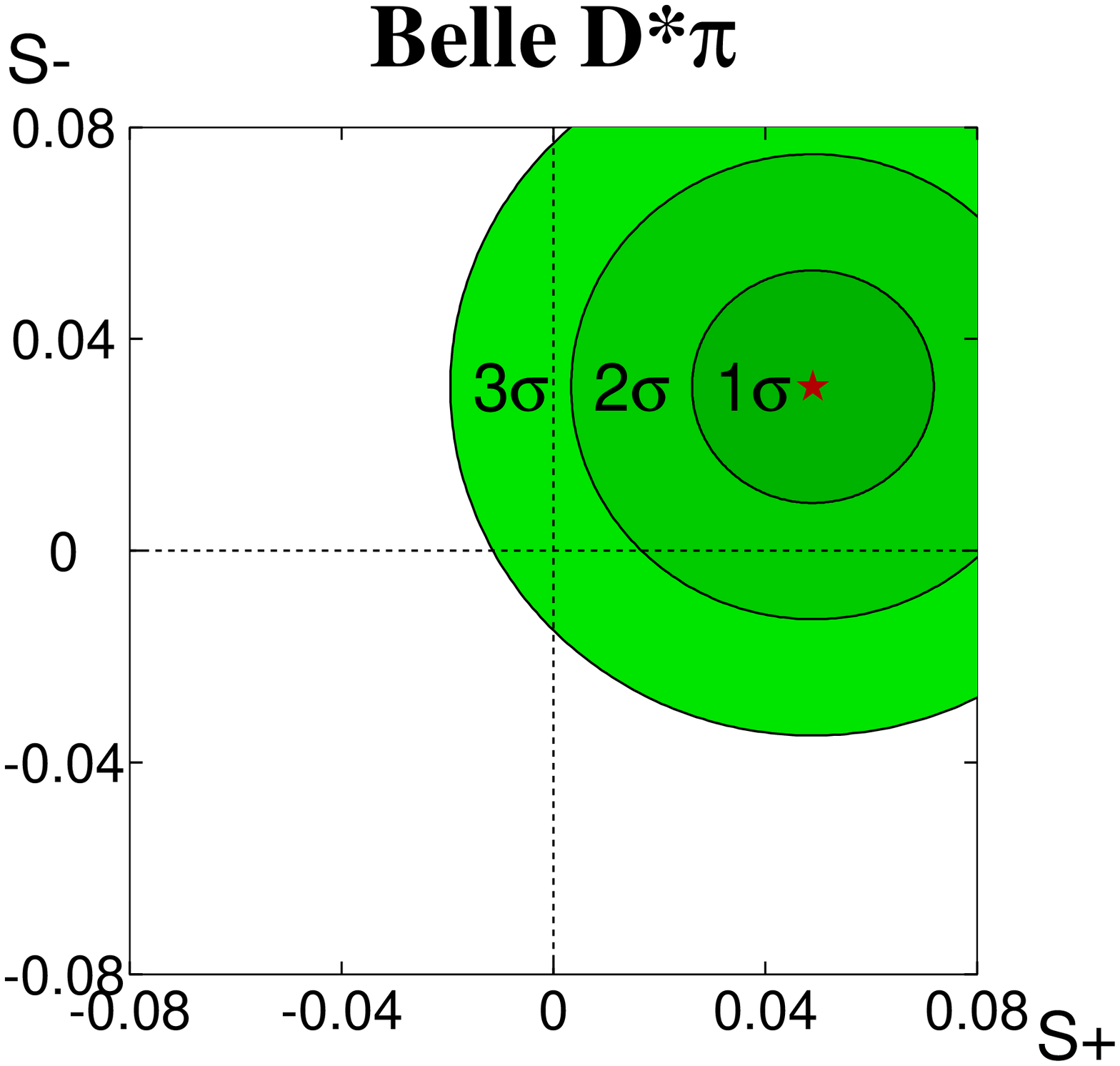}
    \includegraphics[width=0.235\textwidth]{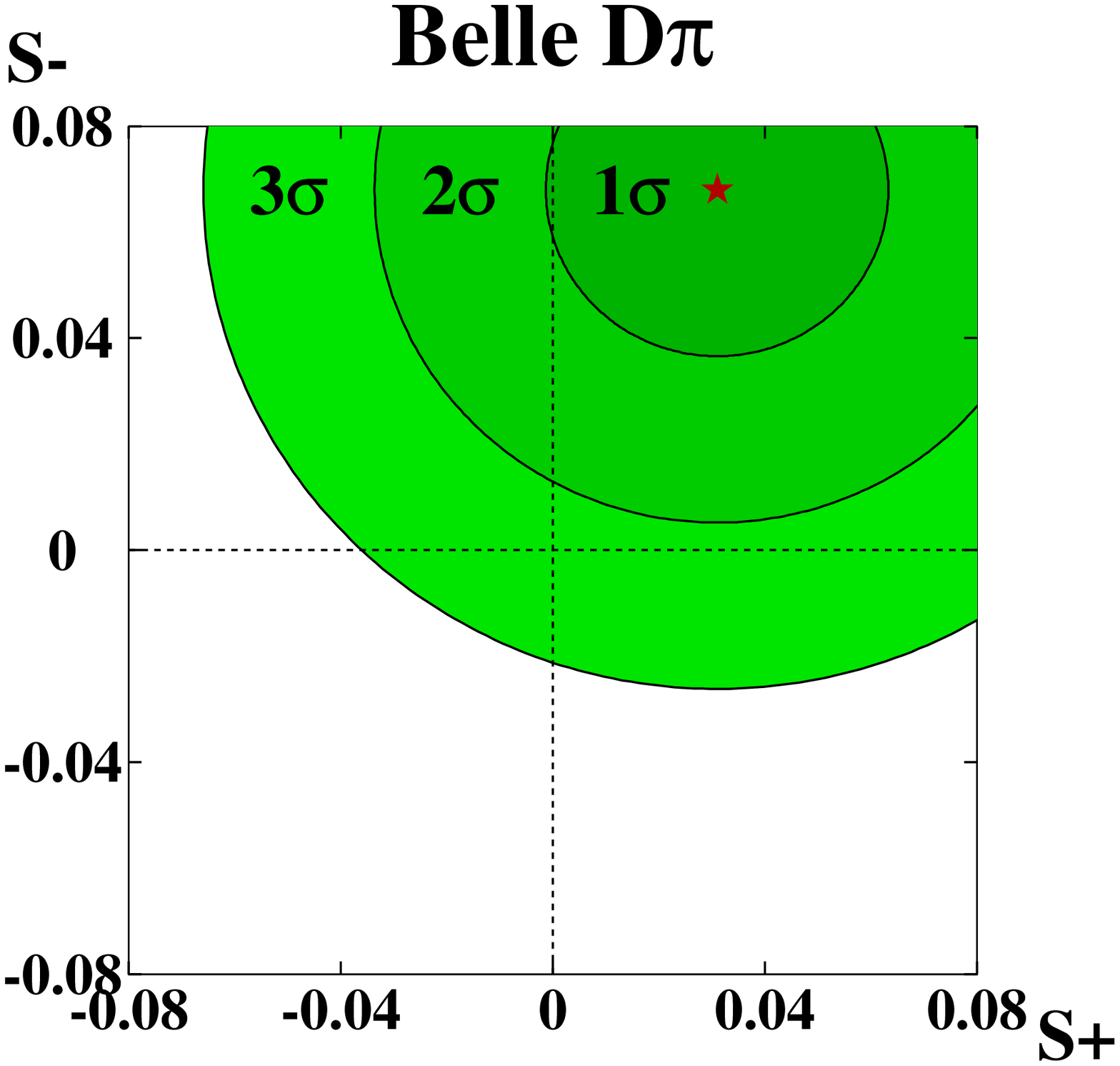}
  \end{center}
  \caption{Results of the $S^\pm$ measurements expressed in terms of
 $S^-$ vs $S^+$ for the $D^* \pi$ (left) and $D \pi$ (right) modes.
 Shaded regions indicate  allowed regions with $1\,\sigma$, $2\,\sigma$
 and $3\,\sigma$ uncertainties defined
 by $\sqrt{-2\mathrm{ln}{\cal L}} = 1,\, 4,\, 9$, respectively 
}
    \label{fig:s-final}
\end{figure}

The $D^* \pi$ signal candidates in the full reconstruction sample and
the partial reconstruction sample are mostly independent. We find only 60
$D^* \pi$ candidates which enter both samples. We repeat the analysis
of the partial reconstruction sample after removing those overlapping
candidates and obtain the same result. 
Therefore we combine the two results. 
Some part of the systematic errors in the two measurements may be
correlated. We conservatively assume that contributions from physics
parameters and fit biases (fit and MC bias in the full reconstruction
sample and $\Delta z$ offsets in the partial reconstruction sample) 
in the two measurements are 100\% correlated, and combine them 
by taking a weighted average using the inverse of the statistical errors 
as weights. 
The final results are
\begin{eqnarray}
S^+ (D^* \pi)&=& 0.049 \pm 0.020 \pm 0.011, \nonumber \\
S^- (D^* \pi)&=& 0.031 \pm 0.019 \pm 0.011, \nonumber \\
S^+ (D \pi)&=& 0.031 \pm 0.030 \pm 0.012, \nonumber \\
S^- (D \pi)&=& 0.068 \pm 0.029 \pm 0.012 
\end{eqnarray}
where the first errors are statistical and the second errors are
systematic. 
These results are shown in Fig.~\ref{fig:s-final} in terms of 
$1\,\sigma$, $2\,\sigma$ and $3\,\sigma$ allowed regions in the $S^-$ versus 
$S^+$ space.
Small possible correlations in the systematic errors of $S^+$ and $S^-$
are neglected. 

The results for $D^* \pi$ and $D\pi$ decay modes show that  
the $(S^+ + S^-)$ values are $2.5\,\sigma$ and $2.2\,\sigma$ away from
zero while the $(S^+ - S^-)$ values are within $1\,\sigma$ of zero.    
Since
$(S^+ + S^-) \propto \sin (2\phi_1+\phi_3) \cos \delta$ and 
$(S^+ - S^-) \propto \cos (2\phi_1+\phi_3) \sin \delta$ (Eq.~\ref{eq:spm}), 
it can be seen that these results are consistent with
both $\delta_{D^* \pi}$ and $\delta_{D\pi}$ being small, 
as predicted by some theoretical models~\cite{wolfenstein}. 
The significance of $CP$ violation, 
seen as deviations of $(S^+ + S^-)$ from zero,
is $2.5\,\sigma$ for $D^* \pi$ and $2.2\,\sigma$ for $D \pi$ decay modes.

\subsection{\boldmath Constraints on $(2\phi_1 +\phi_3)$ and $R$}
\begin{figure}[htb]
  \begin{center}
    \includegraphics[width=0.235\textwidth]{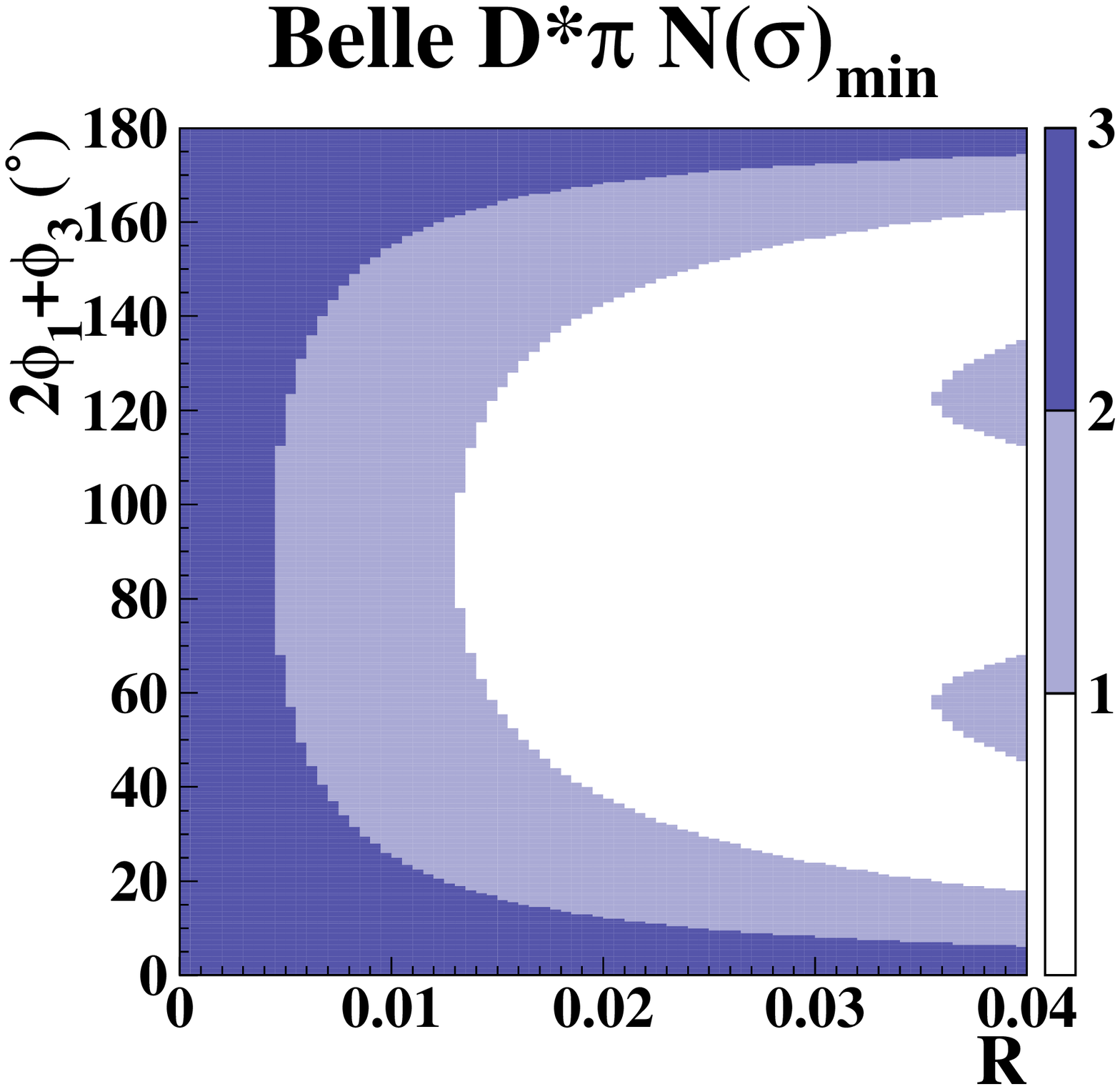}
    \includegraphics[width=0.235\textwidth]{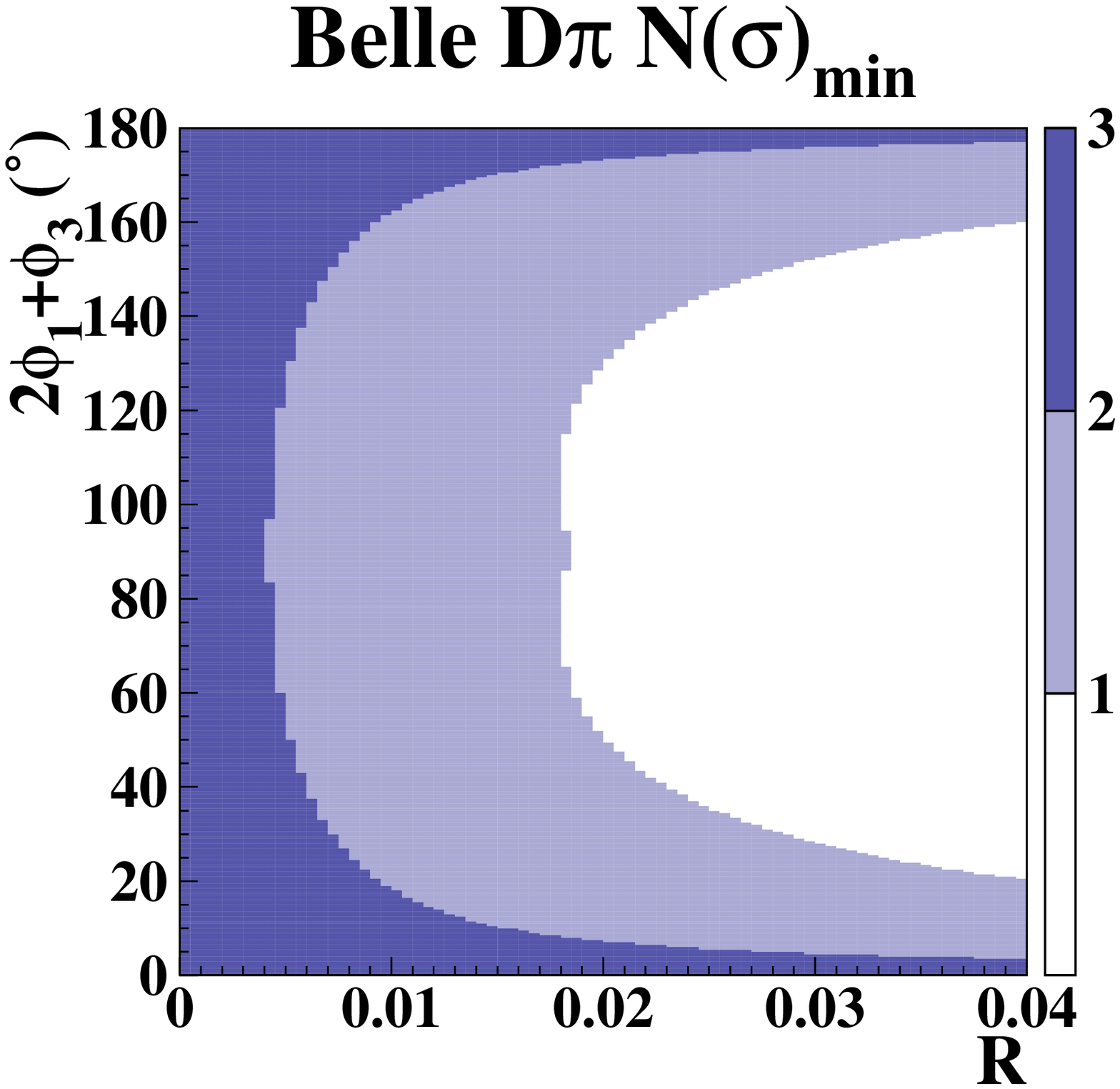}
  \end{center}
  \caption{
    Excluded regions of($2\phi_1+\phi_3$) vs $R$ space at
    $1,~2~,3\,\sigma$ level for the $D^* \pi$ (left) and $D \pi$ (right) decays. 
    The range $0^\circ$--$180^\circ$ for $2\phi_1+\phi_3$ is shown;
    there are additional solutions at 
     $2\phi_1+\phi_3 \longrightarrow 2\phi_1+\phi_3+180^\circ$.
  }
  \label{fig:phasevsr}
\end{figure}

\begin{figure}[htb]
  \begin{center}
    \includegraphics[width=0.235\textwidth]{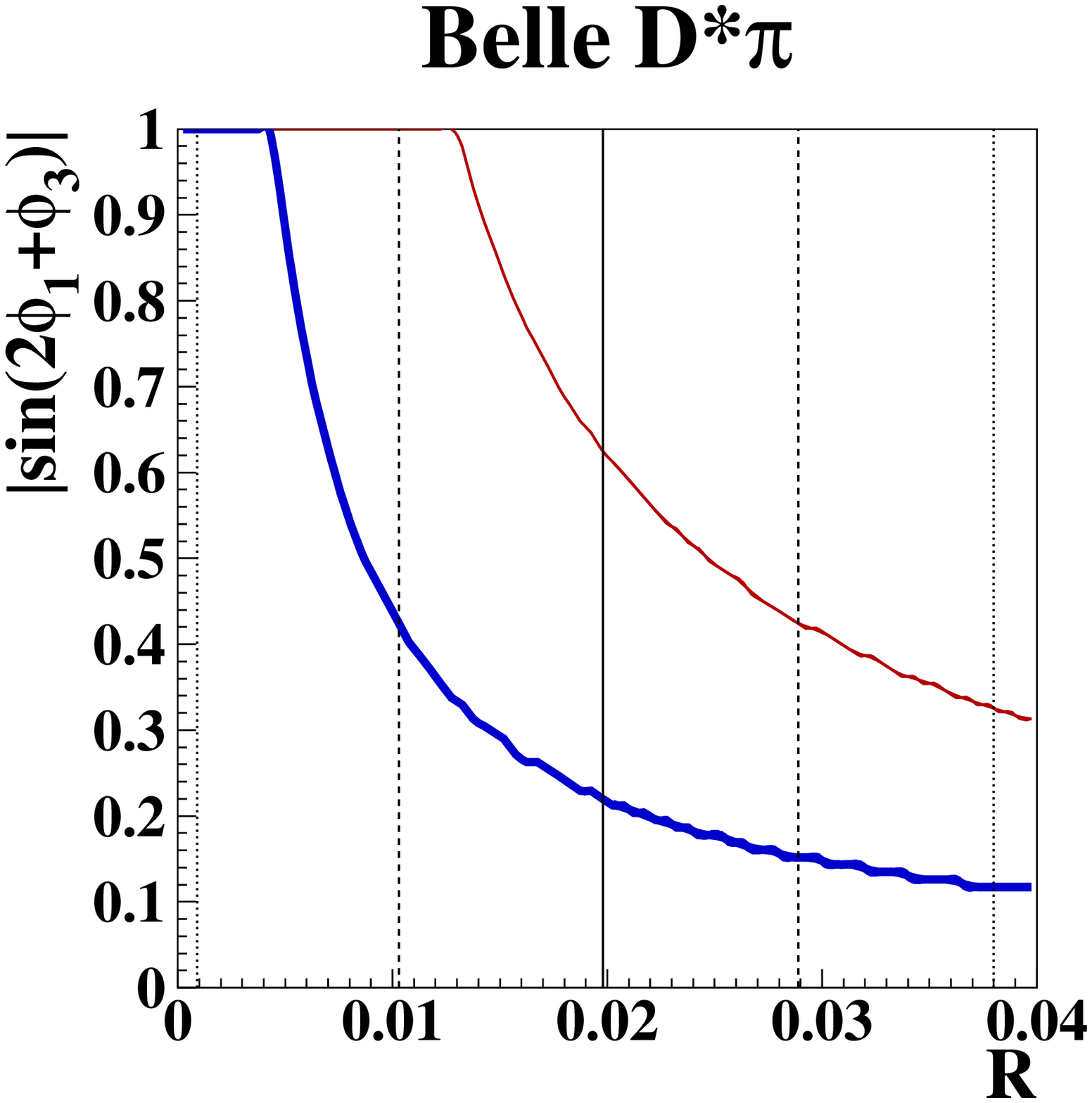}
    \includegraphics[width=0.235\textwidth]{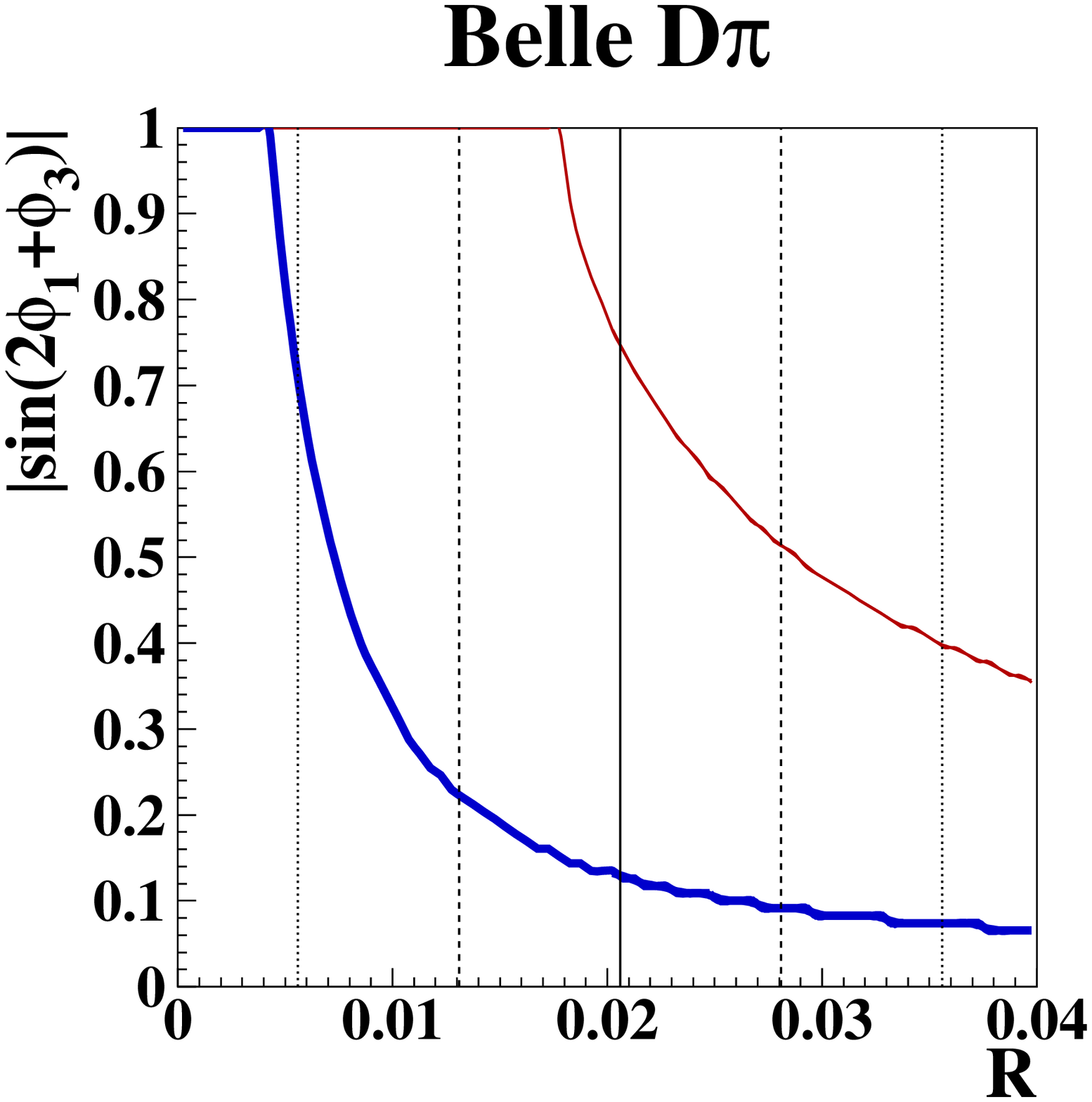}
  \end{center}
  \caption{
    Lower limit on $|\sin(2\phi_1+\phi_3)|$ as a function of $R$ 
    at $1\sigma$ (68\% CL) (thick curves) and $2\sigma$ (95\% CL)
 (thin curves) from (left) $D^* \pi$ and (right) $D\pi$. 
  The estimated values for $R_{D^{(*)}\pi}$ are indicated as solid lines
 (central values). Coarse dotted lines and fine dotted lines indicate 
  their $1\sigma$ and $2\sigma$ errors. 
  }
  \label{fig:phi}
\end{figure}

Since we have two measurements ($S^+$ and $S^-$) which depend on three
unknowns ($R$, $2\phi_1 +\phi_3$, $\delta$), there is not sufficient
information to solve for the weak phase $(2\phi_1 +\phi_3)$. 
Instead we obtain exclusion regions in two-dimensional space 
for any value of the third variable. 
The regions of $(2\phi_1 +\phi_3,~R)$ space that are excluded at
the $1,~2,~3\sigma$ levels are shown in Fig.~\ref{fig:phasevsr}. 
An alternative representation,
shown in Fig.~\ref{fig:phi}, gives the lower bound on 
$|\sin(2\phi_1+\phi_3)|$ for any values of $R$ and $\delta$.

Further conclusions cannot be drawn without some theoretical 
estimate of the values of either $R_{D^{(*)}\pi}$ or $\delta_{D^{(*)}\pi}$.
One interesting possibility is to estimate the size of $R^2_{D^{(*)}\pi}$
using decays such as $B^+ \to D^{(*)\pm} \pi^0$ and $B^0 \to D_s^{(*)\pm} \pi^\mp$,
which are related to $B^0 \to D^{(*)\pm} \pi^\mp$ by 
isospin and SU(3), respectively~\cite{dunietz}.

The method using SU(3) symmetry has some experimental advantages, 
since the rates are enhanced by the square of the tangent of the Cabibbo 
angle $\theta_c$. The relevant expression is
\begin{equation}
R^2_{D^{(*)}\pi} = \tan^2 \theta_c
 \left(\frac{f_{D^{(*)}}}{f_{D_s^{(*)}}}\right)^2
     \frac{{\cal B}(B^0 \to D_s^{(*)+} \pi^-)}
          {{\cal B}(B^0 \to D^{(*)-} \pi^+)},
\end{equation}
where $f_M$ is the decay constant
for the $M$-meson.
The equality is valid up to SU(3) breaking effects.
Both Belle~\cite{belle_dspi} and BaBar~\cite{babar_dspi} have reported
evidence for the decay $B^0 \to D_s^+ \pi^-$. 
BaBar included a limit for the $B^0 \to D_s^{*+}\pi^-$ mode in the same
paper.
We estimate $R_{D^* \pi}$ and $R_{D \pi}$ using the measured branching
fractions and the form factors from a lattice QCD calculation 
by the UKQCD collaboration~\cite{ukqcd}. 
We use  
${\cal B}(B^0 \to D_s^{*+} \pi^-)/{\cal B}(B^0 \to D^{*-} \pi^+)= 0.0068^{+0.0047}_{-0.0051}$, 
which is determined from the BaBar result 
${\cal B}(B^0 \to D_s^{*+}\pi^-) = (1.9^{+1.2}_{-1.3}\pm 0.5) \times 10^{-5}$ 
and the PDG 2004 value ${\cal B}(B^0 \to D^{*-} \pi^+) = (2.76 \pm 0.21) \times 10^{-3}$,
and 
${\cal B}(B^0 \to D_s^{+} \pi^-)/{\cal B}(B^0 \to D^{-} \pi^+)= 0.0098\pm0.0040$, 
which is determined from the PDG 2004 values of 
${\cal B}(B^0 \to D_s^{+} \pi^-) = (2.7 \pm 1.0) \times 10^{-5}$ 
and 
${\cal B}(B^0 \to D^{-} \pi^+) = (2.76 \pm 0.25) \times 10^{-3}$. 
The decay constant ratios are 
$f_{D^*}/f_{D_s^*} =1.04 \pm 0.01  \pm 0.02$ 
and 
$f_{D}/f_{D_s} = 0.90 \pm 0.01  \pm 0.01$.  
We use $\tan \theta_c = 0.23$~\cite{PDG,thetac}.

Adding $\pm 30\%$ theory uncertainty for SU(3) breaking effects, 
such as contributions from annihilation diagrams,
we obtain 
\begin{eqnarray}
  R_{D^* \pi} & = & 0.020 \pm 0.007 \pm 0.006 (\mathrm{theory}), 
  \nonumber \\ 
  R_{D \pi} & = & 0.021 \pm 0.004 \pm 0.006 (\mathrm{theory})
\end{eqnarray}
as indicated in Fig.~\ref{fig:phi}. 
If we use these values, we can
restrict the allowed region of $|\sin (2\phi_1 + \phi_3)|$  to be 
\begin{equation}
  |\sin (2\phi_1 + \phi_3)|> 0.44~(0.13)~~\mathrm{at}~68\%(95\%)\,\mathrm{CL}
\end{equation}
from the results for the $D^*\pi$ mode and 
\begin{equation}
  |\sin (2\phi_1 + \phi_3)|> 0.52~(0.07)~~~\mathrm{at}~68\%(95\%)\,\mathrm{CL}
\end{equation}
from the results for the $D\pi$ mode, respectively. 

\section{Summary}
 
We have measured $CP$ violation parameters that depend on $\phi_3$ using the 
time-dependent decay rates of the $B^0 \to D^{(*)\mp} \pi^\pm$.  
A total of 386 million $B \overline{B}$ events were used in the analysis. 
While the $D\pi$ sample was collected by a standard full reconstruction
method, the $D^*\pi$ sample was enlarged by using, in addition, a partial
reconstruction technique,  essentially doubling the statistical power of
this mode. 

The final results expressed in terms of $S^+$ and $S^-$, which are
related to the CKM angles $\phi_1$ and $\phi_3$, the ratio of suppressed to 
favoured amplitudes, and the strong phase difference between them,  as 
$S^{\pm} = -R_{D^*\pi} \sin(2\phi_1+\phi_3 \pm \delta_{D^*\pi})/
                \left( 1 + R_{D^*\pi}^2 \right)$ for $D^* \pi$ and 
$S^{\pm} = +R_{D\pi} \sin(2\phi_1+\phi_3 \pm \delta_{D\pi})/
                \left( 1 + R_{D\pi}^2 \right)$ for $D \pi$,
are 
\begin{eqnarray}
S^+ (D^* \pi)&=& 0.049 \pm 0.020 \pm 0.011, \nonumber \\
S^- (D^* \pi)&=& 0.031 \pm 0.019 \pm 0.011, \nonumber \\
S^+ (D \pi)&=& 0.031 \pm 0.030 \pm 0.012, \nonumber \\
S^- (D \pi)&=& 0.068 \pm 0.029 \pm 0.012, 
\end{eqnarray}
where the first errors are statistical and the second errors are systematic.
These results are an indication of $CP$ violation in  
$B^0 \to D^{*-}\pi^+$ and $B^0 \to D^- \pi^+$ decays at the 
$2.5\,\sigma$ and $2.2\,\sigma$ levels, respectively.  
If we use the values of  $R_{D^* \pi}$ and $ R_{D \pi}$ that are
determined using a combination of factorization and SU(3)
symmetry assumptions, the branching fraction measurements for the 
$D_s^{(*)} \pi$
modes, and lattice QCD calculations, we obtain 68\% (95\%) confidence
level lower limits on $|\sin (2\phi_1 + \phi_3)|$ of 
0.44 (0.13) and 0.52 (0.07) from the $D^* \pi$ and $D \pi$ modes, 
respectively.

\section*{Acknowledgements}
We thank the KEKB group for the excellent operation of the
accelerator, the KEK cryogenics group for the efficient
operation of the solenoid, and the KEK computer group and
the National Institute of Informatics for valuable computing
and Super-SINET network support. We acknowledge support from
the Ministry of Education, Culture, Sports, Science, and
Technology of Japan and the Japan Society for the Promotion
of Science; the Australian Research Council and the
Australian Department of Education, Science and Training;
the National Science Foundation of China under contract
No.~10175071; the Department of Science and Technology of
India; the BK21 program of the Ministry of Education of
Korea and the CHEP SRC program of the Korea Science and
Engineering Foundation; the Polish State Committee for
Scientific Research under contract No.~2P03B 01324; the
Ministry of Science and Technology of the Russian Federation; 
the Ministry of Higher Education, 
Science and Technology of the Republic of Slovenia;  
the Swiss National Science Foundation; 
the National Science Council and the Ministry of Education of Taiwan; 
and the U.S. Department of Energy.


\begin{thebibliography}{99}
\bibitem{dzb}
  A small amount of data is reprocessed using a set of constants which
  is corrected for the small relative misalignment of the SVD and CDC.
  The correction function is calculated from comparisons 
  of the vertex positions for tracks between the two reprocessings.
\end{thebibliography}
\end{document}